\documentclass[12pt]{article}
\usepackage{graphicx} 
\usepackage{jheparxiv}
\usepackage{hyperref}
\hypersetup{
    colorlinks,
    citecolor=black,
    filecolor=black,
    linkcolor=black,
    urlcolor=black
}
\usepackage[english]{babel}
\usepackage{tikz}
\usetikzlibrary{positioning,arrows.meta,calc}
\usepackage{tikz-cd}
\usepackage{tensor}
\usepackage{amsfonts}
\usepackage{amsmath}
\usepackage{amssymb}
\usepackage{amsthm}
\usepackage{mathtools}
\usepackage{tensor}
\usepackage{slashed}
\usepackage{mathrsfs}
\usepackage{bm}
\usepackage{setspace}
\usepackage{amssymb}
\usepackage{dsfont}
\usepackage[T1]{fontenc}
\usepackage{cancel}
\usepackage{mdframed}
\input{twistor.sty}

\usepackage{bbm}
\usepackage{graphicx}
\usepackage{stmaryrd}
\usepackage{subcaption}
\usepackage{cancel}
\usepackage{bigints}

\usepackage{bbm}
\usepackage{graphicx}
\usepackage{stmaryrd}
\usepackage{subcaption}

\usepackage{cancel}
\usepackage{bigints}

\title{
	AdS$_4$ Celestial Symmetries: from Ambitwistor charges 
to CFT$_3$ Light-Ray Operators}

\author[a]{Alex Goodenbour,}
\author[a]{Adam Kmec,}
\author[a]{Lionel Mason,}
\author[b,c,d]{Romain Ruzziconi}

\affiliation[a]{The Mathematical Institute, University of Oxford,\\ Woodstock Road, Oxford OX2 6GG, United
Kingdom\vspace{0.1cm}}
\emailAdd{alexander.goodenbour@maths.ox.ac.uk}
\emailAdd{adam.kmec@maths.ox.ac.uk}
\emailAdd{lionel.mason@maths.ox.ac.uk}

\affiliation[b]{Center for the Fundamental Laws of Nature, Harvard University \\
17 Oxford Street, Cambridge, MA 02138, USA \vspace{0.1cm}}

\affiliation[c]{Black Hole Initiative, Harvard University \\
20 Garden Street, Cambridge, MA 02138, USA \vspace{0.1cm}}

\affiliation[d]{Centre de Physique Théorique, École polytechnique, CNRS\\
91128 Palaiseau Cedex, France \vspace{0.1cm}}

\emailAdd{romainruzziconi@fas.harvard.edu}

\abstract{The ${L}_\Lambda w_{1+\infty}$ algebra is the cosmological-constant deformation of the celestial $Lw_{1+\infty}$ symmetry algebra of gravity, providing an infinite-dimensional extension of the three-dimensional conformal algebra at the AdS$_4$ boundary. We construct the Hamiltonian charges associated with these symmetries from boundary ambitwistor-space data, using LeBrun's Heaven on Earth correspondence, and translate them to spacetime via the Penrose transform. We relate the resulting expressions to light-ray operators in the boundary CFT$_3$, including ANEC operators, their conformal descendants, and their commutators.
}

\begin{document}
\maketitle

\section{Introduction}

Celestial symmetry algebras connect soft limits, asymptotic charges and light-ray operators to  geometric gauge transformations of asymptotic twistor spaces. For Yang--Mills theory and gravity with flat asymptotics, the relevant algebras are the $S$-algebra and $Lw_{1+\infty}$. They were obtained from soft collinear limits and celestial operator products \cite{Strominger:2021lvk,Guevara:2021abz}.  Twistor space provides their geometric interpretation as Hamiltonian or gauge transformations \cite{Adamo:2021lrv,Bu:2022iak,Mason:2022hly}, extending the original self-dual constructions \cite{Penrose:1976jq,Penrose:1976js,Ward:1977ta}.
They were subsequently obtained from top-down constructions \cite{Costello:2022wso,Costello:2023hmi,Bittleston:2024efo}. Their spacetime realizations include charges at null infinity \cite{Freidel:2021ytz,Freidel:2023gue,Cresto:2024fhd,Cresto:2024mne,Cresto:2025bfo,Cresto:2025ubl} with their twistor origins given in \cite{,Kmec:2024nmu} and on finite-distance null hypersurfaces \cite{Ruzziconi:2025fct,Ruzziconi:2025fuy,Kmec:2026dis,Freidel:2026iia}.

A priori these various approaches are quite distinct,  but a first step toward relating the spacetime and twistor perspectives was taken for gravity in \cite{Kmec:2024nmu}. There the canonical charges were derived from first principles by applying covariant phase-space methods to the twistor-space action \cite{Mason:2007ct} and were then expressed in terms of spacetime data at null infinity using the Penrose transform. The connection with twisted and celestial holography was subsequently developed for the $S$-algebra of Yang--Mills theory in \cite{Kmec:2025ftx}, where the Carrollian and celestial realizations of the charges were shown to arise from different choices of Cauchy surface in twistor space and at null infinity.

A nonzero cosmological constant deforms the gravitational algebra to $L_\Lambda w_{1+\infty}$ \cite{Taylor:2023ajd,Bittleston:2024rqe} with the latter work expressing the algebra as Poisson diffeomorphisms of the AdS$_4$ twistor space with the generators $w^p_{m,a}$ simply expressed in terms of Hamiltonians given as homogeneous monomials in the standard twistor coordinates on twistor space as in \eqref{eq:wmonomials}. The $L_\Lambda w_{1+\infty}$ commutators read as 
\begin{equation}
    \begin{split}
   \left\{w^p_{m,a},\,w^{p'}_{m',a'}\right\}
    &= \left(m(p'-1)-m'(p-1)\right)w^{p+p'-2}_{m+m',a+a'} \\
    &\qquad\qquad\qquad\qquad\qquad    - \frac{\Lambda}{6}\left(a'(p-2)-a(p'-2)\right)w^{p+p'-1}_{m+m',a+a'} ,
\end{split}  \label{algebra LLambda}
\end{equation} where the range of indices for the generators $w^p_{m,a}$ is $p\in\frac{1}{2}\mathbb{Z}$, $m\pm p \in \mathbb{Z}$, $a \pm p \in \mathbb{Z}$, $p+m-1 \geq 0$,   $a-p +2\geq 0$. Recent work has also realized this algebra through light-ray operators in conformal field theories \cite{Himwich:2025ekg,Himwich:2026exq,Sheta:2025oep,Strominger:2026yrh,Alday:2026rso,Zhu:2026ova}. In particular, the CFT$_3$ construction uses the averaged null energy operator (ANEC), its conformal descendants, and their commutators \cite{Strominger:2026yrh}.

The goal of the present paper is to provide a spacetime dictionary between that light-ray operator construction and the boundary representatives of the twistor charges. In particular, we relate the twistor construction of the ${L}_\Lambda w_{1+\infty}$ algebra to charge algebras defined directly in spacetime in terms of charge aspects integrated over two-dimensional surfaces. When such a surface is chosen to be ruled by null geodesics, one of the integrations is naturally performed along the light rays, so that the resulting charge can be expressed as a one-dimensional integral of light-ray operators. In this way, we systematically relate the spacetime charges to light-ray operators in the boundary CFT, including ANEC operators and their conformal descendants as discussed in \cite{Strominger:2026yrh}. Although the symmetries are defined in the non-linear theory, in the translation to spacetime, we work in the simplified setting of linearised metric perturbations around an AdS background. This allows us to construct the charges without imposing a self-duality condition in the bulk. This leads to simple, tractable expressions that capture the essential features of the correspondence with ANEC operators and are compatible with standard conformally flat Dirichlet boundary conditions. Our work complements the $\Lambda$-BMS$_4$ approach \cite{Compere:2019bua,Compere:2020lrt,Fiorucci:2020xto} and related treatments of symmetries and phase space analyses in asymptotically (A)dS gravity \cite{Geiller:2022vto,Bonga:2023eml,Compere:2023ktn,McNees:2024iyu,McNees:2025acf,Campoleoni:2023fug}.

Our construction starts from the twistor representation of ${L}_\Lambda w_{1+\infty}$ as local holomorphic contact diffeomorphisms given in \cite{Bittleston:2024rqe}. We obtain its Noether charges from the $\Lambda \neq 0$ twistor actions of \cite{Mason:2007ct} by applying Noether  methods, and then use the Penrose transform to obtain the corresponding spacetime expressions, with the twistor action leading directly to the standard boundary symplectic structures. The twistor theory of asymptotically AdS$_4$ spacetimes makes use of LeBrun's `Heaven on Earth' construction \cite{LeBrun:1982vjh}, which identifies the twistor space of a bulk self-dual asymptotically AdS$_4$ spacetime with the space of complex null geodesics---the \emph{ambitwistor space}---of its conformal boundary $\scri$. We extend the charge formulae derived from the twistor-space action in \cite{Kmec:2024nmu} to the $\Lambda \neq 0$ case, together with boundary ambitwistor-space data adapted to nonzero cosmological constant. A particular feature of \cite{Kmec:2025ftx} is that the algebra and the corresponding charge aspects are built recursively from the underlying spacetime symmetries. This recursion originates from recursion operators associated with the self-duality equations viewed as an integrable system. For Yang--Mills theory, this structure was identified in \cite{Chau:1982mj,MasonWoodhouse:1996,Ivanova:1997cu}, and for gravity in \cite{Boyer:1985aj,Strachan:1993pq,Strachan:1994ts,Dunajski:2000iq}.

We summarize below the key formulae and structures obtained in this paper. First, the recursion relations defining the celestial charge aspects in AdS are given by
\begin{equation}
    \partial_\perp\mathcal{Q}_s
    =
    \partial_+\mathcal{Q}_{s-1}
    +
    \partial_-\mathcal{Q}_{s+1},
    \qquad
    s\in \mathbb{Z}
\label{recursion relations intro}
\end{equation}
where $(y_\perp, y_+, y_-)$ denote AdS boundary coordinates. The charge aspects $\mathcal{Q}_s$ for $|s|\le 2$ parametrize the components of the holographic stress tensor (as constructed from the Fefferman-Graham expansion of the metric around $\scri$), while those with $|s|>2$ are constructed nonlocally from the leading data and subleading terms in the Fefferman--Graham expansion of the metric via this recursion relation. 
The spacetime surface charges associated with the $L_\Lambda w_{1+\infty}$ algebra are of the form
\begin{equation}
    H_\xi[\Sigma]
    =
    \int_\Sigma \d\Sigma^a\,\mathcal J_{\xi a}.
\label{eq:general spacetime charge_intro}
\end{equation}
Here, $\Sigma$ is a two-dimensional surface at the boundary, and $\mathcal J_{\xi a}$ is a conserved boundary current.  The charge aspects  are  encoded in  \eqref{eq:chargeaspecttwistor}  in terms of a homogeneous twistor function $g(p_a,y_a)$ satisfying $p\cdot\p_y g=0$. This also gives the currents  via  the integral formula
\begin{equation}
    \mathcal J_{\xi a}(y)
    :=
    \frac{1}{2\pi i}
    \int_{L_y}
    \D\lambda\wedge
    p_a\,\xi\,
    g\vert_{L_y}.
\label{eq:general boundary current_intro}
\end{equation}
This is a complex integral on the twistor line $L_y \simeq \mathbb{CP}^1$, where $g$ encodes the gravitational data on ambitwistor space, $p_a = (p_\perp, p_+, p_-)$ are momentum coordinates on ambitwistor space with $p_ap^a=0$, and $\xi$ is a symmetry generator, which can be expanded in terms of the $w^p_{m,a}$ entering \eqref{algebra LLambda}. This boundary current is the building block of our discussion and can also be integrated over a null two surface generated by light rays along $\partial_+$ leading to the constituent integrals:
\begin{equation} \label{light-ray}
      \mathcal L_{\xi,+}(y^\perp,y^-)
    :=
    \int_{-\infty}^{+\infty}dy^+\,
    \mathcal J_{\xi+}(y^\perp,y^+,y^-)
\end{equation}
These CFT light-ray operators can be expressed in the Einstein cylinder frame \cite{Sheta:2025oep, Strominger:2026yrh}, where the commutator algebra of their modes corresponds to the $L_\Lambda w_{1+\infty}$ algebra. In particular, taking ${\xi}\vert_{L_y} =p_+$, the current reduces to the corresponding holographic stress tensor components, and the light-ray operator reduces to the averaged null energy operator (ANEC),
\begin{equation}
    \mathcal E(y^\perp)
    =
    \int_{-\infty}^{+\infty}dy^+\,
    T_{++}(y^\perp,y^+,y^-=0).
\label{eq:ANEC-flat-complex_intro}
\end{equation}
This operator is the starting point of the symmetry construction on the Einstein cylinder. The other light-ray operators in \eqref{light-ray} can also be obtained by taking conformal descendants of ANEC and their commutators, generating the $L_\Lambda w_{1+\infty}$ algebra \cite{Sheta:2025oep, Strominger:2026yrh}. In this work, we provide explicit semiclassical expressions for these operators in terms of the celestial charge aspects entering \eqref{recursion relations intro}.

The rest of the paper is organized as follows. In Section \ref{sec:Asymptotically AdS spacetimes}, we review asymptotically AdS spacetimes in Fefferman--Graham gauge and discuss the resulting asymptotic expansions. In particular, we introduce celestial charge aspects in AdS and relate them to the radial expansion of the metric. In Section \ref{sec:twistors}, we introduce the twistor space of the Poincaré patch and the ambitwistor space of its conformal boundary. We then explain the Heaven on Earth correspondence, which relates the bulk twistor space to the boundary ambitwistor space. In Section \ref{ssec:penrose}, we introduce the Penrose transform and use it to reconstruct the holographic stress tensor and higher-spin celestial charge aspects in AdS from ambitwistor-space data. In Section \ref{sec:symmetries}, we explain how the ${L}_\Lambda w_{1+\infty}$ algebra emerges in this set-up and in what sense its generators act as symmetries of the theory. Finally, in Section \ref{sec:Charges}, we compute the charges associated with the ${L}_\Lambda w_{1+\infty}$ algebra and relate them to the ANEC operators and their descendants. Appendix \ref{Appendix: Further Details} contains further details on the twistor theory presented in the main text.

\vspace{0.2cm}

\textit{Note added: While this paper was in preparation, the interesting parallel work  \cite{DiGiacomo:2026oku} appeared  discussing the ${L}_\Lambda w_{1+\infty}$ charges from a spacetime perspective. }

\section{Asymptotically AdS spacetimes}
\label{sec:Asymptotically AdS spacetimes}

\subsection{Linearised gravity in AdS}
\label{sec:Linearized gravity in AdS}

In this section, we review the definition of asymptotically AdS$_4$ spacetimes in linearised gravity \cite{deHaro:2000vlm, Compere:2019bua}. We work in a Poincaré patch with coordinates $ x^\mu = (z, y^\perp , y^+, y^-)$, where $z> 0$ is the holographic coordinate ($z=0$ at the boundary of AdS$_4$) and $x^a = (y^\perp , y^+, y^-)$ are the boundary coordinates. The AdS$_4$ metric takes the following form  
\begin{equation} \label{Poincare metric}
     g^{\mathrm{AdS}} = \frac{\ell^2}{z^2} (dz^2 - (dy^\perp )^2 + dy^+ dy^-)
\end{equation}
and the corresponding conformal boundary metric is 
\begin{equation}
    [ds^2_{\partial \mathrm{AdS}}] = [- (dy^\perp)^2 + dy^+ dy^-]
    \label{conf boundary metric}
\end{equation} Here, for convenience, we are working with split (Kleinian) signature  conventions if   $(y^\perp, y ^+, y^-)$ are taken to be real, with $y^\pm$ null, but Lorentzian and Euclidean signature can be accommodated by allowing the coordinates to be complex, e.g., with $y^+$ the complex conjugate of $y^-$ for Lorentz signature. 

Let us now consider a linearised gravitational perturbation $h_{\mu\nu}$ on the AdS$_4$ background, subject to the Fefferman-Graham \cite{Fefferman:2007rka} gauge conditions
\begin{equation}
    h_{zz} = 0 = h_{za}
\end{equation} The metric $g^{\mathrm{AdS}}_{\mu\nu} + h_{\mu\nu}$ is asymptotically AdS$_4$  provided we have the following expansion  
\begin{equation} \label{sec:dirichlet falloff}
    h_{ab} = \frac{\ell^2}{z^2}\sum_{n=0}^\infty  z^nh^{(n)}_{ab} 
\end{equation} where $h^{(n)}_{ab}$ are functions of $x^a$.  We take homogeneous Dirichlet boundary conditions  so that the conformal boundary metric \eqref{conf boundary metric} is fixed giving $h^{(0)}_{ab}=0$.
Einstein's equations then imply that $h^{(1)}_{ab} = 0 = h^{(2)}_{ab}$ but  then allow arbitrary choices for $h^{(3)}_{ab}$  as free datum.  All the remaining  orders  with $n>3$ are then completely determined by this datum.

 In the AdS/CFT correspondence, $h^{(3)}_{ab}$ is identified with the VEV of the dual CFT stress tensor, which we will denote as\footnote{\label{fn:stress-normalization}In the usual AdS/CFT conventions \cite{deHaro:2000vlm}, the VEV of the stress tensor is given by $\langle \mathcal{T}_{ab}  \rangle = \frac{3\ell^2}{16\pi G_{\mathrm N}}\,h_{(3)ab}$. Here, we defined the stress tensor as \eqref{stress tensor} to omit the global factor in front of the charge expressions below. In other words, we have $T_{ab} = - \frac{4\pi G_{\mathrm{N}}}{\ell}\langle \mathcal{T}_{ab}  \rangle $. 
}
\begin{equation}
  \boxed{  T_{ab}
    =
    -\frac{3\ell}{4} \,h_{(3)ab}, }
\label{stress tensor}
\end{equation}
 As a consequence of the linearised Einstein's equations, we have
\begin{equation} \label{stress tensor conditions}
   \boxed{ \partial^a T_{ab} = 0, \qquad {T^a}_a = 0  }
\end{equation} where the index is raised with the boundary metric as expected for a CFT stress tensor. 

\subsection{Weyl spinors and the Newman-Penrose formalism}

We introduce the Weyl tensor and its
corresponding spinor decomposition encoding the self-dual and anti-self-dual parts.
Writing a spacetime index as
$\mu\equiv \alpha\dot\alpha$, the chiral decomposition of the Weyl tensor into its anti-self-dual and self-dual parts is
\begin{equation} \label{weyl tensor in terms of spinors}
    C_{\alpha\dot\alpha\,\beta\dot\beta\,
      \gamma\dot\gamma\,\delta\dot\delta}
    =
    \psi_{\alpha\beta\gamma\delta}\,
    \epsilon_{\dot\alpha\dot\beta}
    \epsilon_{\dot\gamma\dot\delta}
    +
    \overline{\psi}_{\dot\alpha\dot\beta\dot\gamma\dot\delta}\,
    \epsilon_{\alpha\beta}
    \epsilon_{\gamma\delta},
\end{equation}
where $\psi_{\alpha\beta\gamma\delta}=
    \psi_{(\alpha\beta\gamma\delta)}$ and $\overline{\psi}_{\dot\alpha\dot\beta\dot\gamma\dot\delta}
=\overline{\psi}_{(\dot\alpha\dot\beta\dot\gamma\dot\delta)}$ are the Weyl spinors. These are complex conjugate in Lorentzian signature, but independent in other signatures or in the complex. 

At a hypersurface (here $\scri$) with normal $N^\mu=N^{\alpha\dot\alpha}$ the electric and magnetic parts of the Weyl tensor are 
\begin{equation}
E_{\mu\nu}=C_{\mu\lambda\nu\rho}N^\lambda N^\rho, \qquad B_{\mu\nu}=C_{\mu\lambda\sigma\tau}\half \varepsilon^{\sigma\tau}_{\nu\rho}N^\lambda N^\rho .    
\end{equation}
These are orthogonal to $N^\mu$ and so define tensors  $E_{ab}, B_{ab}$ on the hypersurface. 
Once $N^{\alpha\dot\alpha}$ has been normalized so that $N^2=2$, it can be used to convert dotted spinor indices into undotted ones. The 3-component indices $a, b,\ldots $ for vectors orthogonal to $N$ can then be identified with symmetric  pairs of undotted indices $a=(\alpha\beta)$ and so on.  With this,  $E_{ab}$ can be identified with the sum $\psi_{\alpha\beta\gamma\delta}+\bar \psi_{\alpha\beta\gamma\delta}$ and $B_{ab}$ the difference.

In the Fefferman-Graham expansion around $\scri$, the leading contribution to $E_{ab}$ can be identified with $h_{(3)ab}$ and so encodes the holographic stress-energy tensor. In practice, we develop our formulae when the conformal structure of $\scri$ is flat, so that the Cotton tensor vanishes. Since the Cotton tensor determines the leading magnetic part of the Weyl tensor, $B_{ab}=0$ at leading order and the two chiral Weyl spinors are therefore not independent at the boundary. It is thus sufficient to work with one chirality, as we will mostly do in the following.

The Bianchi identities give the zero-rest-mass field equations
\begin{equation} \label{Bianchi linearized}
    \nabla^{\alpha\dot\alpha}
    \psi_{\alpha\beta\gamma\delta}
    =0,
    \qquad
    \nabla^{\alpha\dot\alpha}
    \overline{\psi}_{\dot\alpha\dot\beta\dot\gamma\dot\delta}
    =0.
\end{equation}
Contracting these equations with $N^{\alpha\dot\alpha}$ gives 
\begin{equation}
  D^{\alpha\beta}  \psi_{\alpha\beta\gamma\delta}=0, \qquad D_{\alpha\beta}:=N_{(\alpha}^{\dot\alpha}\nabla_{\beta)\dot\alpha}\, .
\end{equation}
This gives $D^aE_{ab}=0=D^aB_{ab}$. At leading order at $\scri$, identifying the leading part of $E_{ab}$ with the holographic stress-energy tensor yields $D^aT_{ab}=0$.

In spinor indices, the metric convention is
$g_{\alpha\dot{\alpha}\beta\dot{\beta}}
= -\epsilon_{\alpha\beta}\bar{\epsilon}_{\dot{\alpha}\dot{\beta}}$.
Introducing normalized spinor dyads
$(o^\alpha,\iota^\alpha)$ and
$(\bar{o}^{\dot{\alpha}},\bar{\iota}^{\dot{\alpha}})$, with
$o_\alpha\iota^\alpha
=1
=\bar{o}_{\dot{\alpha}}\bar{\iota}^{\dot{\alpha}}$, the five Newman--Penrose scalars $\Psi_i$ associated
with $\psi_{\alpha\beta\gamma\delta}$ are defined by   \cite{Newman:1961qr} 
\begin{align}
    \Psi_0
    &=
    \psi_{\alpha\beta\gamma\delta}
    o^\alpha o^\beta o^\gamma o^\delta,
    &
    \Psi_1
    &=
    \psi_{\alpha\beta\gamma\delta}
    \iota^\alpha o^\beta o^\gamma o^\delta,
    \nonumber
    &\Psi_2
    =
    \psi_{\alpha\beta\gamma\delta}
    \iota^\alpha\iota^\beta o^\gamma o^\delta,
    \\
    \Psi_3
    &=
    \psi_{\alpha\beta\gamma\delta}
    \iota^\alpha\iota^\beta\iota^\gamma o^\delta,
    &
    \Psi_4
    &=
    \psi_{\alpha\beta\gamma\delta}
    \iota^\alpha\iota^\beta\iota^\gamma\iota^\delta.
\end{align}
The dotted Weyl spinor admits an analogous decomposition into $\overline{\Psi}_i$, $i=0,\ldots,4$.

We will define the associated Newman--Penrose tetrad 
\begin{equation}
\begin{aligned}
    l^{\alpha\dot\alpha}
    =o^\alpha\bar o^{\dot\alpha}, \quad
    n^{\alpha\dot\alpha}
    =\iota^\alpha\bar\iota^{\dot\alpha}, \quad 
    m^{\alpha\dot\alpha}
=o^\alpha\bar\iota^{\dot\alpha}, \quad 
    \bar m^{\alpha\dot\alpha}
    =\iota^\alpha\bar o^{\dot\alpha}.
\end{aligned}
\end{equation} to be parametrized in terms of the Poincaré coordinates as 
\begin{equation}
\begin{aligned}
    l
    =\frac{z}{\sqrt{2}\,\ell}
      \left(
          \partial_{y^\perp}-\partial_z
      \right), \quad
    n
    =\frac{z}{\sqrt{2}\,\ell}
      \left(
          \partial_{y^\perp}+\partial_z
      \right), \quad 
    m
    =\frac{\sqrt{2}\,z}{\ell}\,
      \partial_{y^+}, \quad
\bar m
    =\frac{\sqrt{2}\,z}{\ell}\,
      \partial_{y^-}.
\end{aligned}
\label{eq:NP-tetrad-Poincare}
\end{equation}
Here $l$ is directed toward the conformal boundary $z=0$, whereas $n$
is directed toward the Poincaré interior. The tetrad satisfies 
\begin{equation}
    l^2=n^2=m^2=\bar m^2=0,
    \qquad
    l\cdot n=-1,
    \qquad
    m\cdot\bar m=1,
\end{equation}
 with all other contractions vanishing. The corresponding one-forms are
\begin{equation}
\begin{aligned}
    l_\mu dx^\mu
    &=-\frac{\ell}{\sqrt{2}\,z}
      \left(dy^\perp+dz\right),
    &
    n_\mu dx^\mu
    &=\frac{\ell}{\sqrt{2}\,z}
      \left(dz-dy^\perp\right),
    \\
    m_\mu dx^\mu
    &=\frac{\ell}{\sqrt{2}\,z}\,dy^-,
    &
    \bar m_\mu dx^\mu
    &=\frac{\ell}{\sqrt{2}\,z}\,dy^+,
\end{aligned}
\end{equation}
so that the metric can be reconstructed as $g_{\mu\nu}^{\mathrm{AdS}}
    =-2l_{(\mu}n_{\nu)}
     +2m_{(\mu}\bar m_{\nu)}$.

For our Dirichlet solutions to the linearised Einstein equations,
the Newman--Penrose components of the Weyl spinor in the
tetrad \eqref{eq:NP-tetrad-Poincare} have the uniform falloff
\begin{equation}
   \boxed{ \Psi_n^{\mathrm{FG}}(z,y)
    =
    \left(\frac{z}{\ell}\right)^3    
        \Psi_{n,\mathrm{FG}}^{(0)}(y)
        +\mathcal{O}\left({z}^4\right),
    \qquad n=0,\ldots,4 . }
\label{eq:FG-peeling}
\end{equation}
We have the same result for $\overline{\Psi}_n$.

To make the analogy with peeling in asymptotically flat space, we  can instead employ a tetrad whose outgoing null vector is affinely parametrized. Introducing
\begin{equation}
    u= \ell \left(y^\perp+z\right),
    \quad
    r=\frac{\ell}{z}, \qquad \mbox{ or }\qquad 
 \label{from FG to Bondi}
    z=\frac{\ell}{r},
    \quad
    y^\perp= \frac{u}{\ell}-\frac{\ell}{r},
\end{equation}
and the AdS$_4$ metric becomes
\begin{equation}
    ds^2_{\mathrm{AdS}}
    =
    -\frac{r^2}{\ell^2}\,du^2
    -2\,du\,dr
    +r^2\,dy^+dy^- .
\end{equation}
A tetrad analogous to an affinely parametrized Newman--Unti \cite{Newman:1962cia} tetrad is
\begin{equation}
\begin{aligned}
    (l_{\mathrm{NU}},  n_{\mathrm{NU}},   m_{\mathrm{NU}} ,    \bar m_{\mathrm{NU}})= \left(\partial_r, \partial_u
      -\frac{r^2}{2\ell^2}\partial_r, \frac{\sqrt{2}}{r}\partial_{y^+},
\frac{\sqrt{2}}{r}\partial_{y^-}\right).
\end{aligned}
\end{equation}
It is related to the regular Fefferman--Graham tetrad by the boost
\begin{equation}
    (l_{\mathrm{NU}},    n_{\mathrm{NU}})=(A\,l_{\mathrm{FG}},
    A^{-1}n_{\mathrm{FG}}),
    \qquad
    A=\sqrt{2} z
     =\sqrt{2}\,\ell\,\frac{1}{r},
\end{equation}
with $m$ and $\bar m$ unchanged. Since $\Psi_n$ has boost weight
$2-n$, one finds
\begin{equation}
    \Psi_n^{\mathrm{NU}}
    =A^{2-n}\Psi_n^{\mathrm{FG}}.
\end{equation}
Using \eqref{eq:FG-peeling}, this gives
\begin{equation}
    \Psi_n^{\mathrm{NU}}
    =
    \left(\sqrt{2}\ell\right)^{2-n}\frac{1}{r^{5-n}}
        \Psi_{n,\mathrm{FG}}^{(0)}
        (y^\perp=u/\ell,y^+,y^-)
        +\mathcal{O}\left(\frac{1}{r^{6-n}}\right)
\label{eq:FG-to-NU-peeling}
\end{equation}
Defining the leading Newman-Unti coefficient by
\begin{equation}
    \Psi_n^{\mathrm{NU}}
    =
    \frac{\Psi_{n,\mathrm{NU}}^{(0)}(u,y^+,y^-)}
         {r^{5-n}}
    +\mathcal{O}\left(r^{n-6}\right),
\label{eq:AdS-NU-peeling}
\end{equation}
one obtains
\begin{equation}
    \boxed{
    \Psi_{n,\mathrm{NU}}^{(0)}(u,y^+,y^-)
    =\left(\sqrt{2}\ell\right)^{2-n}
    \,
    \Psi_{n,\mathrm{FG}}^{(0)}
    (y^\perp=u/\ell,y^+,y^-),
    \qquad n=0,\ldots,4 .
    }
\label{eq:Bondi-FG-Weyl-coefficients}
\end{equation}
The same relation holds for the anti-self-dual coefficients
$\overline{\Psi}_{n,\mathrm{NU}}^{(0)}$ and
$\overline{\Psi}_{n,\mathrm{FG}}^{(0)}$.

\subsection{Celestial charge aspects and conformal symmetries}
\label{sec:Celestial charge aspects and conformal symmetries}

  The radial  Fefferman-Graham expansion gives for the Weyl components
\begin{equation}
    \Psi_{n}^{\mathrm{FG}}(z,y)
    =
    \left(\frac{z}{\ell}\right)^3
    \sum_{m=0}^{\infty}
    \left(\frac{z}{\ell}\right)^m
    \Psi_{n,\mathrm{FG}}^{(m)}(y).
\end{equation} Plugging this expansion into the linearised Bianchi identities \eqref{Bianchi linearized} yields 
\begin{equation}\label{FullBianchiIdentities}
\begin{split}
    \partial_\perp \Psi_{n+1,FG}^{(m)} - 2\partial_- \Psi_{n,FG}^{(m)} - \frac{m+1}{l}\Psi_{n+1,FG}^{(m+1)} &= 0\\
    \partial_\perp \Psi_{n,FG}^{(m)} - 2\partial_+ \Psi_{n+1,FG}^{(m)} + \frac{m+1}{l}\Psi_{n,FG}^{(m+1)} &= 0
\end{split}
\end{equation}
where $m = 0,1,\dots$ is the index for the subleading orders and $n=0,1,2,3$. These two sets of equations have different roles. For
$k=1,2,3$, adding the first equation in
\eqref{FullBianchiIdentities} with $n=k-1$ to the second one with
$n=k$ eliminates the coefficient at order $m+1$ and gives
\begin{equation}
    \partial_\perp\Psi_{k,\mathrm{FG}}^{(m)}
    =
    \partial_+\Psi_{k+1,\mathrm{FG}}^{(m)}
    +
    \partial_-\Psi_{k-1,\mathrm{FG}}^{(m)}.
\label{TangentialBianchiIdentities}
\end{equation}
Taking instead their difference gives the radial recursion
\begin{equation}
    \frac{m+1}{\ell}\Psi_{k,\mathrm{FG}}^{(m+1)}
    =
    \partial_+\Psi_{k+1,\mathrm{FG}}^{(m)}
    -
    \partial_-\Psi_{k-1,\mathrm{FG}}^{(m)}.
\label{RadialBianchiIdentities}
\end{equation}
The two unpaired endpoint equations are
\begin{align}
    \frac{m+1}{\ell}\Psi_{0,\mathrm{FG}}^{(m+1)}
    &=
    2\partial_+\Psi_{1,\mathrm{FG}}^{(m)}
    -
    \partial_\perp\Psi_{0,\mathrm{FG}}^{(m)},
    \label{RadialBianchiEndpointPlus}
    \\
    \frac{m+1}{\ell}\Psi_{4,\mathrm{FG}}^{(m+1)}
    &=
    \partial_\perp\Psi_{4,\mathrm{FG}}^{(m)}
    -
    2\partial_-\Psi_{3,\mathrm{FG}}^{(m)}.
    \label{RadialBianchiEndpointMinus}
\end{align}
Thus, at each value of $m$, the eight equations in
\eqref{FullBianchiIdentities} decompose into three tangential
constraints and five equations determining the next radial
coefficients. In particular, they do not give two independent
tangential hierarchies as they do in the asymptotically flat case.

We now  introduce the notation for the celestial
charge aspects
\begin{equation}
    \mathcal{Q}_s
    :=
    \Psi_{2-s,\mathrm{FG}}^{(0)},
    \qquad
    s=-2,\ldots,2 .
\label{charge aspects}
\end{equation} In terms of those, Equation \eqref{TangentialBianchiIdentities} at $m=0$ then gives the recursion relations 
\begin{equation}
    \boxed{
    \partial_\perp\mathcal{Q}_s
    =
    \partial_+\mathcal{Q}_{s-1}
    +
    \partial_-\mathcal{Q}_{s+1},
    \qquad
    s=-1,0,1 .
    }
\label{recursion relations}
\end{equation}
Thus, the three aspects $\mathcal{Q}_{-1}$, $\mathcal{Q}_0$, and
$\mathcal{Q}_1$ evolve in $y^\perp$ driven by the two extremal-helicity data
$\mathcal{Q}_{-2}$ and $\mathcal{Q}_2$.

To compare with the usual Newman--Unti form of the AdS Bianchi
identities, let
\begin{equation}
    \mathcal{Q}^{\mathrm{NU}}_s
    :=
    \Psi_{2-s,\mathrm{NU}}^{(0)}
    =
    \left(\sqrt{2}\ell\right)^s\mathcal{Q}_s ,
\end{equation}
where we used \eqref{eq:Bondi-FG-Weyl-coefficients}. Introducing the
planar spin-weighted derivatives
\begin{equation}
    \eth=\sqrt{2}\,\partial_+,
    \qquad
    \bar\eth=\sqrt{2}\,\partial_-,
\end{equation}
equation \eqref{recursion relations} becomes
\begin{equation}
    \boxed{
    \partial_u\mathcal{Q}^{\mathrm{NU}}_s
    =
    \eth\mathcal{Q}^{\mathrm{NU}}_{s-1}
    -\frac{\Lambda}{6}\,
     \bar\eth\mathcal{Q}^{\mathrm{NU}}_{s+1},
    \qquad
    s=-1,0,1 ,
    }
\label{Bondi recursion relations}
\end{equation}
where
\begin{equation}
    \Lambda=-\frac{3}{\ell^2}
\end{equation}
is the cosmological constant appearing in
$R_{\mu\nu}=\Lambda g_{\mu\nu}$. This agrees, up to Newman--Penrose
sign and dyad conventions, with the AdS Bianchi identities written in
\cite{Mao:2019ahc,Geiller:2022vto}. Notice that the explicit
cosmological-constant dependence in
\eqref{Bondi recursion relations} is a consequence of the
spin-dependent normalization relating the Bondi and
Fefferman--Graham charge aspects.

The boundary stress tensor \eqref{stress tensor} can be expressed in terms of the
celestial charge aspects as
\begin{equation}
    \boxed{
    T_{ab}
    = \frac{1}{4}
    \begin{pmatrix}
        4\mathcal{Q}_0
        & 2\mathcal{Q}_1
        & 2\mathcal{Q}_{-1}
        \\[2mm]
        2\mathcal{Q}_1
        & \mathcal{Q}_2
        & \mathcal{Q}_0
        \\[2mm]
        2\mathcal{Q}_{-1}
        & \mathcal{Q}_0
        & \mathcal{Q}_{-2}
    \end{pmatrix},
    \qquad
    a,b=(\perp,+,-).
    }
\label{stress tensor charge aspects}
\end{equation}
Indeed, for the boundary metric \eqref{conf boundary metric}, $\eta^{\perp\perp}=-1$, $\eta^{+-}=\eta^{-+}=2$, and
the trace vanishes identically.
Furthermore, the three conservation equations in \eqref{stress tensor conditions} are precisely equivalent to the three recursion relations
\eqref{recursion relations}. Thus, the leading linearised bulk
Bianchi identities are mapped to the tracelessness and conservation
of the holographic stress tensor.

Every conformal Killing vector $\xi^a$ of the boundary metric \eqref{conf boundary metric}
defines a conserved current
\begin{equation}
    J_\xi^a=T^a{}_b\xi^b,
    \qquad
    \partial_aJ_\xi^a=0. \label{conserved current}
\end{equation}
Indeed, using the conformal Killing equation and the tracelessness of
the stress tensor,
\begin{equation}
    \partial_aJ_\xi^a
    =
    T^{ab}\partial_{(a}\xi_{b)}
    =
    \frac{1}{3}\,
    (\partial_c\xi^c)\,T^a{}_a
    =0.
\end{equation}

The conformal charge can be evaluated on any codimension-one
hypersurface $\Sigma$,
\begin{equation} \label{eq:conformal charge}
 \boxed{  H_\xi[\Sigma]
    =
    \int_\Sigma d\Sigma_a\,T^a{}_b\xi^b. }
\end{equation}
We shall consider the three natural choices of $\Sigma$:
\begin{equation}
    \Sigma_\perp:=\{
    y^\perp=\mathrm{const.}\}\, , \qquad 
    \Sigma_-:=\{
    y^-=\mathrm{const.}\}\, , \qquad 
    \Sigma_+:=\{
    y^+=\mathrm{const.}\}.
\end{equation}
Restricting $d\Sigma_a=\frac{1}{2}\epsilon_{abc}dy^b\wedge dy^c$
 appropriately, these will have corresponding oriented area elements 
  \begin{equation}
     d\Sigma^\perp_a    =
    -\delta_a^\perp\,d^2y,\qquad    d\Sigma^-_a
    =
    \frac{1}{2}\delta_a^-\,
    dy^\perp\,dy^+,
\qquad    d\Sigma_a^+ 
    =
    \frac{1}{2}\delta_a^+\,
    dy^\perp\,dy^-,
\end{equation}
where $d^2y=\frac{1}{2}dy^+dy^-$.  Using
$\eta^{\perp\perp}=-1$ etc, the corresponding conformal charges are
\begin{align}
    H_\xi [\Sigma_\perp]
    &=
    \int_{\Sigma_\perp}d\Sigma_a\,T^a{}_b\xi^b =
    \int_{\Sigma_\perp}d^2y\,
    T_{\perp b}\xi^b
    \nonumber\\
    &=
    \int_{\Sigma_\perp}d^2y
    \left(
        T_{\perp\perp}\xi^\perp
        +T_{\perp+}\xi^+
        +T_{\perp-}\xi^-
    \right).
\label{eq:conformal-charge-perp-T}
\end{align}
Using \eqref{stress tensor charge aspects}, this becomes
\begin{equation}
    \boxed{
    H_\xi[\Sigma_\perp]
    =
    \frac{1}{4} \int_{\Sigma_\perp}d^2y
    \left(
        4\mathcal{Q}_0\xi^\perp
        +2\mathcal{Q}_1\xi^+
        +2\mathcal{Q}_{-1}\xi^-
    \right).
    }
\label{eq:conformal-charge-perp-Q}
\end{equation}
Similarly
\begin{align}
    H_\xi [\Sigma_-]
    &=
     \frac{1}{4}\int_{\Sigma_-}dy^\perp\,dy^+
    \left(
        2\mathcal{Q}_1\xi^\perp
        +\mathcal{Q}_2\xi^+
        +\mathcal{Q}_0\xi^-
    \right),
\label{eq:conformal-charge-minus-Q}
\\
    H_\xi [\Sigma_+]
    &=
     \frac{1}{4}\int_{\Sigma_+}dy^\perp\,dy^-
    \left(
        2\mathcal{Q}_{-1}\xi^\perp
        +\mathcal{Q}_0\xi^+
        +\mathcal{Q}_{-2}\xi^-
    \right).
   \label{eq:conformal-charge-plus-Q}
\end{align}
The conservation \eqref{conserved current} implies that these three expressions are different representations of the same
covariant conformal charge assuming that the surfaces can be continuously deformed into each other, so that
\begin{equation}
    H_\xi [\Sigma_\perp]
    =
    H_\xi [\Sigma_-]
    =
    H_\xi [\Sigma_+].
\end{equation}
For noncompact integration surfaces, this equality requires the
corresponding weighted boundary fluxes to vanish. Reversing the
orientation of any hypersurface reverses the overall sign of the
associated charge.

The ten boundary conformal Killing vectors generating $\mathfrak{so}(3,2)$ can be written explicitly as follows. The translation generators are
\begin{equation}
    P_\perp=\partial_\perp,
    \qquad
    P_+=\partial_+,
    \qquad
    P_-=\partial_- .
\end{equation}
The Lorentz generators may be chosen as
\begin{equation}
    J
    =
    y^+\partial_+-y^-\partial_-,
    \qquad  
    B_\pm
    =
    y^\pm\partial_\perp+2y^\perp\partial_\mp,
\end{equation}
The dilatation generator is
\begin{equation}
    D
    =
    y^\perp\partial_\perp
    +y^+\partial_+
    +y^-\partial_- .
\end{equation}
Finally, the special conformal generators are
\begin{align}
    K_\perp
&    =
    -\left((y^\perp)^2+y^+y^-\right)\partial_\perp
    -2y^\perp y^+\partial_+
    -2y^\perp y^-\partial_-,
 \nonumber \\   
    K_\pm &
    =
    y^\perp y^\mp\partial_\perp
    +(y^\perp)^2\partial_\pm
    +(y^\mp)^2\partial_\mp.
\end{align}
Thus, a general global conformal Killing vector is
\begin{equation}
\begin{aligned}
    \xi
    ={}&
    a^\perp P_\perp+a^+P_++a^-P_-
    +\omega J+\beta^+B_++\beta^-B_-
    \\
    &+\lambda D
    +b^\perp K_\perp+b^+K_++b^-K_- .
\end{aligned}
\label{eq:general-boundary-CKV}
\end{equation}

\subsection{Higher-order celestial charge hierarchy}
\label{sec:Higher-order celestial charge hierarchy}

We now wish to extend the hierarchy of recursion relations \eqref{recursion relations} to $|s|\geq2$ by requiring the same recursion
relation to hold,
\begin{equation}
    \partial_\perp\mathcal Q_s
    =
    \partial_+\mathcal Q_{s-1}
    +
    \partial_-\mathcal Q_{s+1},
    \qquad
    s \in \mathbb{Z}.
\label{HigherRecursionRelations}
\end{equation}
For $s=-1,0,1$, this equation follows directly from the leading
Bianchi identities, whereas for $|s|\geq2$ it recursively defines the
higher charge aspects. The latter will be shown later to correspond to the ${L}_\Lambda w_{1+\infty}$ charge aspects.

For $s=2$, Equation \eqref{HigherRecursionRelations} reads
\begin{equation}
    \partial_\perp\mathcal Q_2
    =
    \partial_+\mathcal Q_1
    +
    \partial_-\mathcal Q_3.
\end{equation}
On the other hand, the second equation in
\eqref{FullBianchiIdentities} with $m=0$ and $n=0$ gives
\begin{equation}
    \partial_\perp\mathcal Q_2
    =
    2\partial_+\mathcal Q_1
    -
    \frac{1}{\ell}\Psi_{0,\mathrm{FG}}^{(1)}.
\end{equation}
Comparing these two equations yields
\begin{equation}
    \mathcal Q_3
    =
    \partial_-^{-1}
    \left(
        \partial_+\mathcal Q_1
        -
        \frac{1}{\ell}\Psi_{0,\mathrm{FG}}^{(1)}
    \right).
\label{Q3HigherHierarchy}
\end{equation}

The next element of the hierarchy is defined by
\begin{equation}
    \partial_\perp\mathcal Q_3
    =
    \partial_+\mathcal Q_2
    +
    \partial_-\mathcal Q_4.
\end{equation}
Substituting \eqref{Q3HigherHierarchy} gives
\begin{equation}
\begin{aligned}
    \mathcal Q_4
    =
    \partial_-^{-2}
    \bigg(
        \partial_\perp\partial_+\mathcal Q_1
        -\frac{1}{\ell}
            \partial_\perp\Psi_{0,\mathrm{FG}}^{(1)}
        -\partial_+\partial_-\mathcal Q_2
    \bigg).
\end{aligned}
\label{Q4Intermediate}
\end{equation}
The Bianchi identities at $(m,n)=(1,0)$ and the radial relation
\eqref{RadialBianchiIdentities} at $(m,k)=(0,1)$ imply
\begin{align}
    \partial_\perp\Psi_{0,\mathrm{FG}}^{(1)}
    &=
    2\partial_+\Psi_{1,\mathrm{FG}}^{(1)}
    -\frac{2}{\ell}\Psi_{0,\mathrm{FG}}^{(2)},
    \\
    \frac{1}{\ell}\Psi_{1,\mathrm{FG}}^{(1)}
    &=
    \partial_+\mathcal Q_0
    -\partial_-\mathcal Q_2.
\end{align}
Equation \eqref{Q4Intermediate} therefore becomes
\begin{equation}
    \mathcal Q_4
    =
    \partial_-^{-2}
    \left(
        \frac{2}{\ell^2}\Psi_{0,\mathrm{FG}}^{(2)}
        +\partial_\perp\partial_+\mathcal Q_1
        -2\partial_+^2\mathcal Q_0
        +\partial_+\partial_-\mathcal Q_2
    \right).
\label{Q4HigherHierarchy}
\end{equation}
Using the leading recursion relation for $\mathcal Q_1$, this can
equivalently be written as
\begin{equation}
    \mathcal Q_4
    =
    \partial_-^{-2}
    \left(
        \frac{2}{\ell^2}\Psi_{0,\mathrm{FG}}^{(2)}
        -\partial_+^2\mathcal Q_0
        +2\partial_+\partial_-\mathcal Q_2
    \right).
\end{equation}
More generally, the positive hierarchy is recursively defined by
\begin{equation}
    \boxed{
    \mathcal Q_{s+1}
    =
    \partial_-^{-1}
    \left(
        \partial_\perp\mathcal Q_s
        -
        \partial_+\mathcal Q_{s-1}
    \right),
    \qquad
    s\geq2.
    }
\end{equation}
Repeated use of the radial Bianchi identities then expresses each
$\mathcal Q_s$ in terms of the leading charge aspects and the
subleading coefficients of $\Psi_{0,\mathrm{FG}}$.

The inverse derivative $\partial_-^{-1}$ requires a choice of Green's
function, or boundary condition.
Consequently, the recursion determines
$\mathcal Q_s$ only up to elements of $\ker\partial_-$ unless suitable
boundary conditions are imposed. In the following, we assume that
such a prescription has been fixed.

An analogous discussion determines the charge aspects with $s<-2$.
For $s=-2$, Equation \eqref{HigherRecursionRelations} reads
\begin{equation}
    \partial_\perp\mathcal Q_{-2}
    =
    \partial_+\mathcal Q_{-3}
    +
    \partial_-\mathcal Q_{-1}.
\end{equation}
On the other hand, the first equation in
\eqref{FullBianchiIdentities} with $m=0$ and $n=3$ gives
\begin{equation}
    \partial_\perp\mathcal Q_{-2}
    =
    2\partial_-\mathcal Q_{-1}
    +
    \frac{1}{\ell}\Psi_{4,\mathrm{FG}}^{(1)}.
\end{equation}
Comparing these two equations yields
\begin{equation}
    \mathcal Q_{-3}
    =
    \partial_+^{-1}
    \left(
        \partial_-\mathcal Q_{-1}
        +
        \frac{1}{\ell}\Psi_{4,\mathrm{FG}}^{(1)}
    \right).
\label{QMinus3HigherHierarchy}
\end{equation}
The next element of the negative hierarchy is defined by
\begin{equation}
    \partial_\perp\mathcal Q_{-3}
    =
    \partial_+\mathcal Q_{-4}
    +
    \partial_-\mathcal Q_{-2}.
\end{equation}
Substituting \eqref{QMinus3HigherHierarchy} gives
\begin{equation}
\begin{aligned}
    \mathcal Q_{-4}
    =
    \partial_+^{-2}
    \bigg(
        \partial_\perp\partial_-\mathcal Q_{-1}
        +\frac{1}{\ell}
            \partial_\perp\Psi_{4,\mathrm{FG}}^{(1)}
        -\partial_+\partial_-\mathcal Q_{-2}
    \bigg).
\end{aligned}
\label{QMinus4Intermediate}
\end{equation}
The Bianchi identities at $(m,n)=(1,3)$ and the radial relation
\eqref{RadialBianchiIdentities} at $(m,k)=(0,3)$ imply
\begin{align}
    \partial_\perp\Psi_{4,\mathrm{FG}}^{(1)}
    &=
    2\partial_-\Psi_{3,\mathrm{FG}}^{(1)}
    +\frac{2}{\ell}\Psi_{4,\mathrm{FG}}^{(2)},
    \\
    \frac{1}{\ell}\Psi_{3,\mathrm{FG}}^{(1)}
    &=
    \partial_+\mathcal Q_{-2}
    -\partial_-\mathcal Q_0.
\end{align}
Equation \eqref{QMinus4Intermediate} therefore becomes
\begin{equation}
    \mathcal Q_{-4}
    =
    \partial_+^{-2}
    \left(
        \frac{2}{\ell^2}\Psi_{4,\mathrm{FG}}^{(2)}
        +\partial_\perp\partial_-\mathcal Q_{-1}
        -2\partial_-^2\mathcal Q_0
        +\partial_+\partial_-\mathcal Q_{-2}
    \right).
\label{QMinus4HigherHierarchy}
\end{equation}
Using the leading recursion relation for $\mathcal Q_{-1}$, this can
equivalently be written as
\begin{equation}
    \mathcal Q_{-4}
    =
    \partial_+^{-2}
    \left(
        \frac{2}{\ell^2}\Psi_{4,\mathrm{FG}}^{(2)}
        -\partial_-^2\mathcal Q_0
        +2\partial_+\partial_-\mathcal Q_{-2}
    \right).
\end{equation}
More generally, the negative hierarchy is recursively defined by
\begin{equation}
    \boxed{
    \mathcal Q_{s-1}
    =
    \partial_+^{-1}
    \left(
        \partial_\perp\mathcal Q_s
        -
        \partial_-\mathcal Q_{s+1}
    \right),
    \qquad
    s\leq-2.
    }
\end{equation}
Repeated use of the radial Bianchi identities then expresses each
$\mathcal Q_s$ with $s<-2$ in terms of the leading charge aspects and
the subleading coefficients of $\Psi_{4,\mathrm{FG}}$. As in the positive hierarchy, the inverse derivative
$\partial_+^{-1}$ requires a choice of Green's function.

The hierarchy makes manifest the distinguished role of the two
extremal-helicity aspects $\mathcal Q_{\pm2}$. The leading tangential
Bianchi identities determine the $y^\perp$ evolution of
$\mathcal Q_{-1}$, $\mathcal Q_0$, and $\mathcal Q_1$, given
$\mathcal Q_{\pm2}$ and suitable initial data, while the two endpoint
identities determine subleading radial coefficients rather than
constraining the boundary evolution of $\mathcal Q_{\pm2}$. Thus,
$\mathcal Q_{\pm2}$ represent the two independent local functional
degrees of freedom of the conserved traceless boundary stress tensor.
Once boundary conditions and the zero-mode prescriptions for
$\partial_\pm^{-1}$ are fixed, all charge aspects with $|s|>2$ are
recursively generated descendants and introduce no additional local
degrees of freedom.

\section{Twistors and ambitwistors in AdS}
\label{sec:twistors}

Section \ref{ssec:bulktwistors} reviews the twistor space of AdS$_4$ in the Poincaré patch.
Section \ref{ssec:fromboundary} defines ambitwistor space as the space of null geodesics of the boundary metric. Section \ref{sec:From boundary ambitwistor space to bulk twistor space} establishes the equivalence between the bulk twistor space and the boundary ambitwistor space. This remarkable result is an instance of LeBrun's \textit{Heaven on Earth} construction \cite{LeBrun:1982vjh}.

\subsection{Twistor space of AdS\texorpdfstring{$_4$}{4}}
\label{ssec:bulktwistors}

We first arrange the four Poincaré coordinates $(z, y^\perp, y^+, y^-)$ into a $2\times2$ matrix,
\begin{equation}
\label{eq:xmatrix}
    x^{\dot\alpha \alpha} = \frac{1}{\sqrt{2}}\begin{pmatrix} y^- & y^\perp-z \\ y^\perp+z & y^+ \end{pmatrix}
\end{equation}
where the indices are (dotted and undotted) 2-component spinor indices.
The conformal metric is then the determinant,
\begin{equation}
\label{eq:detdx}
    2\det\left(\d x^{\dot\alpha \alpha}\right) = \d z^2 - (\d y^\perp )^2 + \d y^+\d y^-
    = \frac{z^2}{\ell^2}\,g^{\text{AdS}} .
\end{equation}
Thus a  vector \(v^a = v^{\dot{\alpha}\alpha}\) is null exactly
when its matrix has vanishing determinant. If nontrivial, the matrix then has rank one and 
factorises as $v^{\dot\alpha\alpha} = a^{\dot\alpha}b^\alpha$ into 2-spinors.

The twistor space of AdS\(_4\) in the Poincaré patch is
\begin{equation}\label{def twistor}
\boxed{\PT \equiv \CP^3 - \CP^1_{\lambda_{\alpha} = 0} }
\end{equation} with homogeneous coordinates
$Z^A = (\mu^{\dot\alpha},\lambda_\alpha)$. The line $\CP^1_{\lambda_\alpha=0}$ corresponds to the point at infinity of the Poincaré patch. In general spacetime points correspond to lines in twistor space  by the incidence relation
\begin{equation}
\label{eq:incidence}
    \mu^{\dot\alpha} = x^{\dot\alpha\alpha}\lambda_\alpha .
\end{equation}
This can be read in either direction. Holding $x$ fixed and letting $\lambda_\alpha$ range
over its $\CP^1$ gives a linear embedding of that $\CP^1$ in $\CP^3$, giving the line $L_x \subset \PT$ corresponding to $x$ in spacetime. Holding $Z$ fixed instead and asking which $x$ satisfy
\eqref{eq:incidence} gives two linear conditions on four unknowns, hence a two-plane in the complex (which can be real in split signature). Any two
points of that plane differ by a $\delta x^{\dot\alpha\beta}$ with
$\delta x^{\dot\alpha\beta}\lambda_\beta = 0$, which forces
$\delta x^{\dot\alpha\beta} = c^{\dot\alpha}\lambda^\beta$; this has rank one and is therefore
null. So these are totally-null 2-planes that are ASD and conventionally  called
$\beta$-planes ($\alpha$-planes are self-dual).

The geometry of the correspondence depends only on the conformal class $[g^{\text{AdS}}]$. To encode a metric within the conformal class we introduce  the holomorphic one-form 
\begin{align}
\label{eq:tauinfinity}
    \tau &= I_{AB}Z^A\d Z^B \nonumber \\
    &=\frac{1}{\sqrt{2}\ell} (\lambda_\alpha d\mu^\alpha-\mu^\alpha d\lambda_\alpha)
\end{align}
 on $\PT$ built from a choice of infinity twistor $I_{AB}$, where the second equality defines the specific representation that we will use here (dotted indices have been converted to undotted using $N^{\alpha\dot\alpha}$ as before). For $\Lambda\neq0$ the infinity twistor is
non-degenerate, so $\tau$ is a contact form, and the cosmological constant is fixed by its
normalisation against the holomorphic volume form
$\Omega = \tfrac{1}{6}\epsilon_{ABCD}Z^A\d Z^B\wedge\d Z^C\wedge\d Z^D$ on $\CP^3$,
\begin{equation}
\label{eq:contactnorm}
    \tau\wedge\d\tau = \frac{\Lambda}{3}\,\Omega .
\end{equation}

The restriction of $\tau$  to twistor lines determines the scale; it vanishes on those lines corresponding to points of $\scri$. Since $\tau$
carries weight two under projective rescalings of $Z^A$, its restriction $\tau|_{L_x}$ is a global holomorphic section of
$\Omega^1(2)$ over the compact curve $L_x\cong\CP^1$. But $\Omega^1_{\CP^1} = \cO(-2)$, so this
is a section of the trivial bundle and is therefore constant: $\tau|_{L_x}$ is a multiple of the
canonical weight-two one-form
\begin{equation}
    \D\lambda = \lambda_1\d\lambda_0 - \lambda_0\d\lambda_1
\end{equation}
by a coefficient that cannot depend on $\lambda_\alpha$. Writing that coefficient as
\begin{equation}
\label{eq:fdef}
    \tau|_{L_x} = f(x)\,\D\lambda ,
\end{equation} where
$f$ is a spacetime field of conformal weight 1, and fixing the conformal scale by rescaling $f$ to unity, the Einstein metric in the conformal class
\eqref{eq:detdx} is
\begin{equation}
\label{eq:gfromf}
    g_4 =\frac{2\det\left(\d x^{\dot\alpha\beta}\right) }{f(x)^2}\,,
\end{equation}
the constant of proportionality being fixed by \eqref{eq:contactnorm}.

For AdS$_4$ in the Poincar\'e patch, the infinity twistor is, in the split
$Z^A = (\mu^\alpha,\lambda_\alpha)$,
\begin{equation}
\label{eq:infinitytwistor}
    I_{AB} = \frac{1}{\sqrt{2}\ell}\begin{pmatrix} 0 & -\delta_\alpha^{\ \beta} \\
    \delta^\alpha_{\ \beta} & 0\end{pmatrix} ,
\end{equation}
where a primed index has been converted to an unprimed one using the normal introduced in
Section \ref{ssec:fromboundary} below. It is non-degenerate, and its overall scale is
fixed by \eqref{eq:contactnorm}; we absorb that scale into the constant in \eqref{eq:gfromf} and
take
\begin{equation}
\label{eq:tau}
    \tau = \frac{1}{\sqrt{2}\ell}\left(\lambda_\alpha\d\mu^\alpha - \mu^\alpha\d\lambda_\alpha\right)
    = \frac{1}{\ell}\left(\frac{1}{2}\lambda_0^2\,\d y^- + \lambda_0\lambda_1\,\d y^{\perp} + \frac{1}{2}\lambda_1^2\,\d y^+
      + z\,\D\lambda\right) ,
\end{equation}
which has no $\d z$ component. Along $L_x$ only $\lambda_\alpha$ varies, so
$\d\mu^\alpha = x^{\alpha\beta}\d\lambda_\beta$ and
\begin{equation}
\label{eq:taurestgen}
    \tau|_{L_x} = \frac{1}{\sqrt{2}\ell} x^{\alpha\beta}\left(
\lambda_\alpha\d\lambda_\beta - \lambda_\beta\d\lambda_\alpha\right) = \frac{z}{\ell}\,\D\lambda ,
    \qquad z = \frac{1}{\sqrt{2}}\epsilon_{\beta\alpha}x^{\alpha\beta} .
\end{equation}
The bracket is antisymmetric in $\alpha\beta$, so only the antisymmetric part of
$x^{\alpha\beta}$ survives, which by \eqref{eq:xmatrix} is $z$. Hence, $f = z/\ell$ and
\eqref{eq:gfromf} returns the Poincar\'e patch \eqref{Poincare metric}, with
\eqref{eq:contactnorm} setting $\ell^2 = -3/\Lambda$.

\subsection{Ambitwistor space of the boundary}
\label{ssec:fromboundary}

Now we will briefly forget about the bulk and consider only its three-dimensional conformal boundary at \(z=0\). In Poincaré patch coordinates, that is the flat conformal 3-metric,
\begin{equation}
\label{eq:bdymetric}
    [g] = \left[-(\d y^\perp)^2 + \d y^+\d y^-\right].
\end{equation}

The \textit{Heaven on Earth} construction of  \cite{LeBrun:1982vjh} first complexifies the conformal 3-manifold $\scri\rightarrow \scri^\C$, and then constructs  the space of its  complex null geodesics,  its \textit{ambitwistor space} \(\mathbb{PA}\). In general, 
we can take the cotangent bundle restricted to null covectors $p_a$, $T^*\scri^\C|_{p^2=0}$,  and
quotient by the complex null geodesic flow $p\cdot \nabla$. In Appendix \ref{sec:More on ambitwistor space}, we present more details of this construction on a generic curved 3-manifold. The cotangent bundle comes equipped with a natural holomorphic symplectic
structure $\omega = \d p_a\wedge\d y^a$, where $p_a$ are local coordinates on the fibers of the cotangent bundle. There is a natural one-form, the symplectic potential $\vartheta=p_a\,dy^a$,
satisfying $d\vartheta=\omega$, which descends to the quotient \cite{LeBrun:1982vjh} and projectively defines a contact structure.

For the boundary metric \eqref{eq:bdymetric} at hand this can all be done explicitly. In this case, null geodesics are the straight lines
\begin{equation}
    y^\perp (t) = y^\perp(0) + t\,p^\perp , \qquad y^\pm(t) = y^\pm(0) + t\,p^\pm ,
    \qquad -(p^\perp)^2 + p^+p^- = 0 .
\end{equation}
As in Section \ref{ssec:bulktwistors} the null condition can be parametrised by matrices with vanishing determinant, but here we arrange them into symmetric matrices. This means the two 2-spinor indices must be of the same type. We have,
\begin{equation}
\label{eq:pymatrices}
    p^{\alpha\beta} = \frac{1}{\sqrt{2}}\begin{pmatrix} p^- & p^\perp \\ p^\perp & p^+ \end{pmatrix} ,
    \qquad
    y^{\alpha\beta} = \frac{1}{\sqrt{2}}\begin{pmatrix} y^- & y^\perp \\ y^\perp & y^+ \end{pmatrix} ,
    \qquad
    2\det p^{\alpha\beta} = p^+p^- - (p^\perp)^2 .
\end{equation}
Two pairs $(y,p)$ describe the same geodesic if they differ by sliding the base point along
the geodesic, or by rescaling the tangent vector.
Hence
\begin{equation}
\label{eq:PAquotient}
  \boxed{  \PA = \left\{\left(y^{\alpha\beta},p^{\alpha\beta}\right)\,\big|\,
    \det p^{\alpha\beta} = 0 ,\ p\neq0\right\}\Big/\!\sim ,
    \qquad p\sim s\,p , \quad y\sim y + t\,p , }
\end{equation}
with $s\in\C^*$ and $t\in\C$, the second identification being exactly the quotient along the flow of a null geodesic. 

\subsection{From boundary ambitwistor space to bulk twistor space}
\label{sec:From boundary ambitwistor space to bulk twistor space}

\begin{figure}[t]
    \centering
    \includegraphics[width=\linewidth]{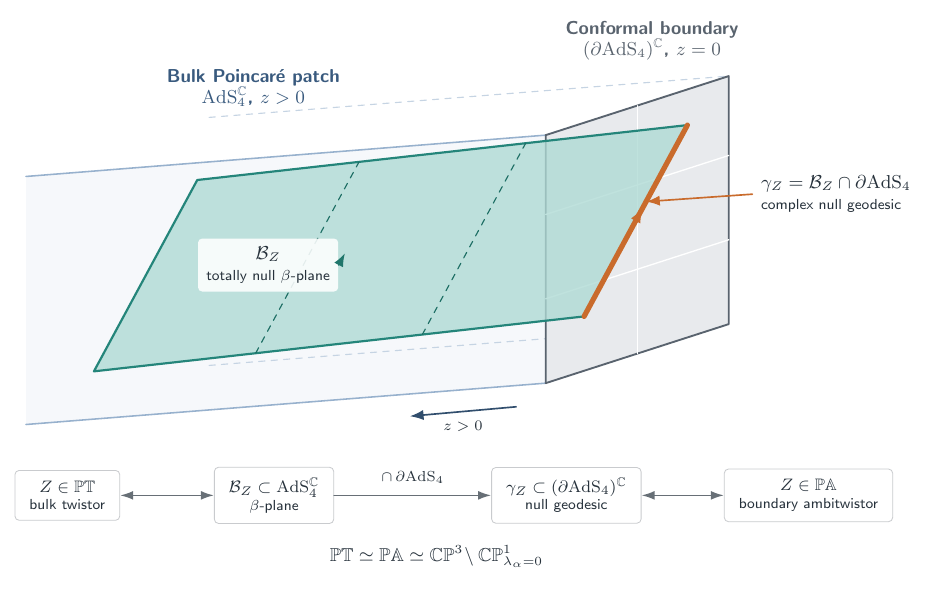}
    \caption{The Heaven on Earth correspondence for the Poincaré patch. A point
$Z\in\mathbb{PT}$ labels a totally null $\beta$-plane $\mathcal B_Z$ in the
complexified bulk. Its intersection with the conformal boundary is a complex
null geodesic $\gamma_Z$, representing the same point $Z\in\mathbb{PA}$ of
boundary ambitwistor space. Hence, $\mathbb{PT}\simeq\mathbb{PA}$.}
    \label{fig:heaven-on-earth}
\end{figure}

The key result of LeBrun's Heaven on Earth construction is that, remarkably, we can identify the bulk twistor space \eqref{def twistor} for a self-dual Einstein spacetime with the  ambitwistor space of its boundary $\scri$ \eqref{eq:PAquotient}. To see this in our flat case,  first note that $p^2=0$ implies 
$\det p_{\alpha\beta} = 0$, and with $p\neq0$ this implies that  the matrix has rank one and is symmetric so that we can write 
\begin{equation}
\label{eq:kernel}
    p_{\alpha\beta}=\frac{1}{\sqrt{2}}\lambda_\alpha\lambda_\beta \, , \quad \mbox{i.e.}\quad (p^-,p^\perp,p^+) = (\lambda_1^2,-\lambda_0\lambda_1,\lambda_0^2)\, .
\end{equation}
For the remaining twistor coordinates we restrict the incidence relations to $\scri$ to set
\begin{equation}
\label{eq:mu}
    \mu^\alpha := y^{\alpha\beta}\lambda_\beta .
\end{equation}
Under $y\mapsto y + tp$ this changes by $t\,p^{\alpha\beta}\lambda_\beta = 0$, so it is
invariant, while under $\lambda\mapsto s\lambda$ it scales as $\mu\mapsto s\mu$. Thus
$(\mu^\alpha,\lambda_\alpha)$ is a point of $\CP^3$, with  $\lambda_\alpha \neq 0$. Conversely,
given such a point, \eqref{eq:mu} is two equations for the three components of the symmetric
matrix $y^{\alpha\beta}$; the map $y\mapsto y^{\alpha\beta}\lambda_\beta$ is onto and its kernel
is spanned by $\lambda^\alpha\lambda^\beta \propto p^{\alpha\beta}$, which is exactly the
ambiguity that we intended to quotient by. Thus we have identified the two spaces.
\begin{equation}
\label{eq:PAiso}
   \boxed{ \PA \;\cong\; \CP^3\setminus\CP^1_{\lambda_\alpha = 0} , }
\end{equation}
with the RHS parametrised by $(\mu^\alpha,\lambda_\alpha)$. This isomorphism is summarized in Figure \ref{fig:heaven-on-earth}.

Thus we have arrived at the twistor space of Section \ref{ssec:bulktwistors}, with
\eqref{eq:mu} as its incidence relation, but with two apparent differences. The index on $\mu$
is undotted here and dotted there; and $y^{\alpha\beta}$ is symmetric, so it has three
components, whereas $x^{\dot\alpha\beta}$ in \eqref{eq:xmatrix} was a general $2\times2$ matrix
carrying four. What is striking is that $\PA$
produces the missing parameter by itself. A boundary point $y$ determines a line parametrised by \(\lambda_{\alpha}\) in \(\mathbb{PA}\) with normal bundle $\cO(1)\oplus\cO(1)$.
Kodaira's theorem then says that the family of such curves is a complex manifold of dimension
$\dim H^0(\CP^1,\cO(1)\oplus\cO(1)) = 4$. So without having even invoked the bulk, the complex geometry tells us there is a fourth dimension that we missed. We can parametrise it as the antisymmetric part of the coordinate matrix,
\begin{equation}
\label{eq:xsplit}
    x^{\alpha\beta} = y^{(\alpha\beta)} -\frac{z}{\sqrt{2}} \,\epsilon^{\alpha\beta} ,
    \qquad z = \frac{1}{\sqrt{2}}\epsilon_{\beta\alpha}x^{\alpha\beta} ,
\end{equation}
the incidence relation $\mu^\alpha = x^{\alpha\beta}\lambda_\beta$ still cuts out a line in
$\CP^3$ for each $x$, and the boundary is the locus $z=0$.

As before, we identify dotted and undotted indices using the normal $N$ to the boundary, normalized so that $N^2=2$ and given by
\begin{equation}
\label{eq:soldering}
N_\alpha{}^{\dot\alpha}
= \iota_\alpha\bar{\iota}^{\dot\alpha}
- o_\alpha\bar{o}^{\dot\alpha}.
\end{equation}
As a vector, $N=\frac{\sqrt{2}\,z}{\ell}\,\partial_z$. Below, we systematically use $N$ to convert dotted indices into undotted ones.

The last piece we need to identify with AdS\(_4\) (as opposed to some other conformally-flat 4-metric) is the conformal scale encoded by the \emph{infinity twistor}. We recall the canonical one-form $\vartheta=p_a\,dy^a$
introduced in Section \ref{ssec:fromboundary}, the symplectic potential on \(\mathbb{PA}\).

In the coordinates of this subsection, the momentum is
$(p^-,p^\perp,p^+) = (\lambda_1^2,-\lambda_0\lambda_1,\lambda_0^2)$, so lowering the index with
$[g]$ gives $p_\perp = \lambda_0\lambda_1$, $p_- = \tfrac12\lambda_0^2$ and
$p_+ = \tfrac12\lambda_1^2$, and the symplectic potential is
\begin{equation}
\label{eq:pdy}
    \vartheta:=p_a\,dy^a
    =\frac12\lambda_0^2\,dy^-
    +\lambda_0\lambda_1\,dy^\perp
    +\frac12\lambda_1^2\,dy^+
    =\frac{1}{\sqrt2}\lambda_\alpha\lambda_\beta\,dy^{\alpha\beta}.
\end{equation}
Now write this in the coordinates \eqref{eq:mu} on $\PA$. Differentiating
$\mu^\alpha = y^{\alpha\beta}\lambda_\beta$,
\begin{equation}
\label{eq:tausplit}
  \frac{1}{\sqrt{2}\ell}\left( \lambda_\alpha\d\mu^\alpha-  \mu^\alpha\d\lambda_\alpha\right)
    =  \frac{1}{\sqrt{2}\ell}y^{\alpha\beta}\left(\lambda_\alpha\d\lambda_\beta - \lambda_\beta\d\lambda_\alpha\right)
      + \frac{1}{\sqrt{2}\ell} \lambda_\alpha\lambda_\beta\,\d y^{\alpha\beta} ,
\end{equation}
and the first term vanishes because $y^{\alpha\beta}$ is symmetric while the bracket is
antisymmetric in $\alpha\beta$. Comparing with \eqref{eq:pdy}, under the boundary incidence relation we obtain
\begin{equation}
\label{eq:tauisPA}
    \tau=\frac{1}{\ell}\,\vartheta
    =\frac{1}{\sqrt2\,\ell}
    \left(\lambda_\alpha\,d\mu^\alpha-\mu^\alpha\,d\lambda_\alpha\right)
    =I_{AB}Z^A\,dZ^B.
\end{equation}
The canonical one-form therefore determines the same contact structure
as the contact form \eqref{eq:tau} of Section \ref{ssec:bulktwistors}. Choosing the normalization
$\tau=\vartheta/\ell$ gives the infinity twistor \eqref{eq:infinitytwistor} and fixes the
bulk AdS radius to $\ell$.

In this construction, points on the boundary are characterised by the vanishing of \(\tau\) when restricted to the corresponding twistor lines,
\begin{equation}
\label{eq:taurest}
    \tau|_{L_x} = \frac{z}{\ell}\,\D\lambda.
\end{equation}
Thus the boundary defining function $z$  is recovered from
twistor data via  $z/\ell = \tau|_{L_x}/\D\lambda$.

We conclude that the \textit{ambitwistor} space of the flat boundary 3-metric (its space of complexified null geodesics) is the twistor space of the bulk AdS\(_4\).

\section{From twistor space to spacetime}
\label{ssec:penrose}

Twistors allow us to solve linear massless field equations on spacetime explicitly in terms of free twistor space data. Here we solve the linearised Bianchi identities \eqref{Bianchi linearized}. We first review the bulk Penrose integral formula,  and then restrict it to the boundary, and this  will allow us to write the celestial charge aspects in terms of this (ambi-) twistor data.

\subsection{Bulk Penrose transform}
\label{sec:Bulk Penrose's transform}

Linearised vacuum gravity can be expressed in terms of the
linearised Weyl tensor \eqref{weyl tensor in terms of spinors}, and the two totally symmetric Weyl
spinors $\psi_{\alpha\beta\gamma\delta}$ and
$\overline\psi_{\dot\alpha\dot\beta\dot\gamma\dot\delta}$ with helicity $-2$ and $+2$ respectively. We focus on the first Bianchi identity in
\eqref{Bianchi linearized}, namely
\begin{equation}
\label{eq:zrm2}
    \nabla^{\alpha\dot\alpha}\psi_{\alpha\beta\gamma\delta} = 0 ,
\end{equation} Equation \eqref{eq:zrm2} is the spin-two case of the
zero-rest-mass equation for a totally symmetric field $\phi_{\alpha_1\cdots\alpha_{2s}}$,
$\nabla^{\alpha_1\dot\alpha}\phi_{\alpha_1\cdots\alpha_{2s}}=0$, whose other cases are the familiar
free equations --- the wave equation at $s=0$, the anti-self-dual half of Maxwell at $s=1$.

The twistor integral formula which solves \eqref{eq:zrm2} is a slight adaptation of the usual Penrose transform,\footnote{The usual spin-two version of the Penrose transform would give rise to a 4-index spinor of conformal weight $-1$, rather than here of weight zero as we use $\tau$ rather than abstract Serre duality on the fibres of the spin bundle over spacetime.} is given by 
\begin{equation} 
\label{eq:penrosespin2}
    \psi_{\alpha\beta\gamma\delta}(x)
    = \frac{1}{2\pi\im}\int_{L_x}
      \tau\vert_{L_x}\wedge\;\lambda_\alpha\lambda_\beta\lambda_\gamma\lambda_\delta\;
      g\big|_{L_x} ,
    \qquad g\in H^1(\PT,\cO(-6) )\, ,
\end{equation}
where $\tau\vert_{L_x} = \frac{z}{\ell}\D \lambda$ is the contact structure restricted to a twistor line $L_x$. Using Dolbeault cohomology, the restriction $g|_{L_x}$ is the $(0,1)$-form $g$ on twistor space restricted to the twistor line $L_x$. This results in the form being a $(0,1)$-form on $\mathbb{CP}^1$ and evaluated on the incidence relation \eqref{eq:incidence},
\begin{equation}
\label{eq:grestrict}
    g\big|_{L_x} := g\!\left(\mu^{\dot\alpha}=x^{\dot\alpha\beta}\lambda_\beta,\lambda_\alpha\right) ,
\end{equation}
and, apart from the explicit factor $z/\ell$, this is the only place $x$ enters: holding $x$ fixed and letting $\lambda_\alpha$ range over its
$\CP^1$ sweeps out the line $L_x$, so the integral is over that line.

The weights then force the rest of the integrand. To integrate over $L_x\cong\CP^1$ we need a
one-form with no projective weight. The canonical measure on the $\CP^1$ is $\D\lambda$, which has
weight $+2$, so if $g$ has homogeneity degree $n$ in $Z^A$ we must supply weight $-n-2$ from
elsewhere. To produce something with four spinor indices we must include four copies of
$\lambda_\alpha$, contributing weight $+4$, so $n=-6$ as in \eqref{eq:penrosespin2}.

The spacetime field depends only on $g\in [g] \in H^{0,1}\!\left(\PT,\cO(-6)\right)$, the Dolbeault cohomology group of sections of $\cO(-6)$, functions of weight $-6$.  This follows from the cohomological gauge freedom in the integral formula \eqref{eq:penrosespin2}: if $g|_{L_x}$ were an exact $(0,1)$-form $g\vert_{L_x} = \bar\partial\vert_{L_x} f$ for some smooth function $f$, then by Stokes' theorem the integral would give the trivial solution $\psi_{\alpha\beta\gamma\delta} = 0$. The closure of $g$ under the action of the Dolbeault operator on $\PT$ results in $\psi_{\alpha\beta\gamma\delta}$ satisfying the z.r.m. equation.

As shown in Appendix \ref{sec:Proof of the Penrose transform in AdS}, differentiating under the integral sign shows that  \eqref{eq:penrosespin2} solves the linearised Einstein Bianchi identity \eqref{eq:zrm2} in AdS$_4$. 

\subsection{Boundary Penrose transform}
\label{sec:Boundary Penrose transform}

We now restrict the Penrose transform \eqref{eq:penrosespin2} to the boundary $z=0$, with incidence relation \eqref{eq:mu}, in \eqref{eq:penrosespin2}.
On the three-dimensional boundary we can identify the two types of spinor indices, and by \eqref{eq:pymatrices} a boundary vector is a symmetric
bispinor, $a\leftrightarrow(\alpha\beta)$. So a totally symmetric spinor of valence four is a trace-free symmetric rank-two tensor.  This is naturally identified with  the CFT$_3$ stress tensor \eqref{stress tensor}. Since
the geodesic momentum is $p_{\alpha\beta}\propto\lambda_\alpha\lambda_\beta$ by \eqref{eq:kernel},
the four factors of $\lambda_\alpha$ in \eqref{eq:penrosespin2} are two factors of $p_a$, and\footnote{We are here invoking Penrose's abstract index notation to identify $a=\alpha\beta$, etc..}
\begin{equation}
\label{eq:Tpenrose}
    T_{ab}(y) \;=\; T_{\alpha\beta\gamma\delta}(y)
    \;=\; \frac{1}{2\pi\im}\int_{L_y}\D\lambda\wedge p_a\,p_b\;g\big|_{L_y} ,
    \qquad L_y = L_x\big|_{z=0} .
\end{equation}
This is one half of the rescaled boundary limit
$\lim_{z\to0}(\ell/z)\psi_{\alpha\beta\gamma\delta}$
of \eqref{eq:penrosespin2}, where the field is expressed in the
conformal spin frame of Appendix \ref{sec:Proof of the Penrose transform in AdS}. Conversion to the physical
spin frame multiplies its components by $(z/\ell)^2$. The same class
$[g]\in H^1(\PT,\cO(-6))=H^1(\PA,\cO(-6))$ generates the bulk ASD Weyl spinor and the boundary
stress tensor.

That it is trace-free follows from contracting under the integral with the boundary metric and using \(p^2 = 0\). What is more surprising is that the conservation condition comes for free as well,
\begin{equation}
\label{eq:Tconserved}
    \partial^a T_{ab}
    = \frac{1}{2\pi\im}\int_{L_y}\D\lambda\wedge p_b\;\left(p^a\partial_a\right)g\big|_{L_y}
    = \frac{1}{2\pi\im}\int_{L_y}\D\lambda\wedge p_b\;p^{\alpha\beta} \lambda_\beta \frac{\partial g}{\partial \mu^\alpha}\big|_{L_y}= 0 ,
\end{equation} where we used the fact that $g|_{L_y}$, being a pullback
from $\PA$, depends only on $y^{\alpha\beta}$ through the incidence relations \eqref{eq:mu}. This is the boundary
avatar of the proof of the bulk Penrose transform given in Appendix \ref{sec:Proof of the Penrose transform in AdS}.

It is useful to strip \eqref{eq:Tpenrose} down to
components. Work in the affine coordinate $q=\lambda_1/\lambda_0$, scale to $\lambda_\alpha=
\lambda_0(1,q)$ so that $\D\lambda=-\lambda_0^2\,\d q$, and let
\begin{equation}
    \hat g(q,y) := -\lambda_0^{6}\,g\big|_{L_y}
\end{equation}
be the weightless trivialisation of $g$. Since each factor $\lambda_\alpha$ contributes either a
$1$ or a $q$, every component of $2T_{\alpha\beta\gamma\delta}$ is of the form  
\begin{equation}
\label{eq:chargeaspecttwistor}
    \boxed{
    \mathcal{Q}_s
    := \frac{1}{2\pi\im}\int_{L_y}q^{\,s+2}\, \d q\wedge \hat g .
    }
\end{equation} In terms of the spinor dyad $(o^\alpha, \iota^\alpha)$ and the five celestial charge aspects \eqref{charge aspects}, we have
\begin{align}
    \mathcal{Q}_{2}
    &=
    2T_{\alpha\beta\gamma\delta}
    o^\alpha o^\beta o^\gamma o^\delta,
    &
    \mathcal{Q}_{1}
    &=
    2T_{\alpha\beta\gamma\delta}
    \iota^\alpha o^\beta o^\gamma o^\delta,
    \nonumber\\
    \mathcal{Q}_{0}
    &=
    2T_{\alpha\beta\gamma\delta}
    \iota^\alpha\iota^\beta o^\gamma o^\delta,
    &
    \mathcal{Q}_{-1}
    &=
    2T_{\alpha\beta\gamma\delta}
    \iota^\alpha\iota^\beta\iota^\gamma o^\delta,
    \nonumber\\
    \mathcal{Q}_{-2}
    &=
    2T_{\alpha\beta\gamma\delta}
    \iota^\alpha\iota^\beta\iota^\gamma\iota^\delta.
\end{align}
Unwinding the dictionary \eqref{eq:pymatrices} between $(\alpha\beta)$ and $a=(\perp,+, -)$, these
are precisely the celestial charge aspects appearing in
\eqref{stress tensor charge aspects}, with $\mathcal{Q}_2\leftrightarrow T_{++}$,
$\mathcal{Q}_{-2}\leftrightarrow T_{--}$ and $\mathcal{Q}_0\leftrightarrow T_{+-}$. 

The recursion relations now follow in one line. Using explicit coordinates,
$p^a \partial_a  g|_{L_y}=0$ says  
\begin{equation}
    \left(\partial_+ - q\,\partial_\perp + q^2\partial_-\right)\hat g = 0 ,
\end{equation}
and multiplying by $q^{\,1+s}$ and integrating, each power of $q$ shifting the label by one,
\begin{equation}
\label{eq:twistorrecursion}
    \boxed{
    \partial_\perp \mathcal{Q}_s
    = \partial_+\mathcal{Q}_{s-1} + \partial_-\mathcal{Q}_{s+1} .
    }
\end{equation}
For $s=-1,0,1$ this is the content of \eqref{eq:Tconserved}, and it reproduces
\eqref{recursion relations}.

In fact the twistor data \(g\) contains more information which we can access if we drop the restriction of $s$ to
$\{-2,\ldots,2\}$. The definition, and the derivation of \eqref{eq:twistorrecursion}, go through
for every $s\in\Z$, and the resulting infinite tower is the hierarchy that Section \ref{sec:Higher-order celestial charge hierarchy} built by hand out of subleading Weyl coefficients: \eqref{eq:twistorrecursion} for
$s=2$ is the relation that there defined $\mathcal{Q}_3$, and so on. What the twistor formula adds
is that the whole tower is read off a single object $g$ on $\PT$, just by varying the power
of $q$.

The $\mathcal{Q}_s$ with $|s|\leq2$ were the components of
$2T_{\alpha\beta\gamma\delta}$. The wider range packages the same way, into higher-spin currents
\begin{equation}
\label{eq:higherspinT}
    \boxed{
    T^{(n)}_{\alpha_1\cdots\alpha_{2n}}(y)
    := \frac{2^{-n/2}}{2\pi\im}\int_{L_y}\D\lambda\wedge
       \frac{\lambda_{\alpha_1}\cdots\lambda_{\alpha_{2n}}}
            {\lambda_0^{\,n-2}\,\lambda_1^{\,n-2}}\;
       g\big|_{L_y} .
    }
\end{equation}
The weights cancel, $+2$ from $\D\lambda$ and $+2n$ from the numerator against $2n-4$ from the
denominator and $-6$ from $g$, and each $\lambda_{\alpha_i}$ contributes a $1$ or a $q$ as before,
so the $2n+1$ components of \eqref{eq:higherspinT} are $2^{-n/2}\mathcal{Q}_n,\ldots,2^{-n/2}\mathcal{Q}_{-n}$. For
$n=2$, the denominator is unity, and we recover \eqref{eq:Tpenrose}. In terms of the spinor dyad $(o^\alpha, \iota^\alpha)$, we have explicitly
\begin{equation}
    \mathcal{Q}_{n-m} = 2^{n/2}\,T^{(n)}_{\alpha_1 \dots\alpha_{2n}} o^{\alpha_1}\cdots o^{\alpha_{2n-m}} \iota^{\alpha_{2n-m+1}}\cdots \iota^{\alpha_{2n}} = \frac{1}{2\pi\im}\int_{L_y} q^{n-m+2}\;\d q\wedge\hat g . 
\end{equation}

Since $2n$ factors of $\lambda_\alpha$ are $n$ factors of $p_a$, \eqref{eq:higherspinT} is a
symmetric rank-$n$ boundary tensor $T^{(n)}_{a_1\cdots a_n}$, trace-free by $p^2=0$ and conserved
by $p^a \partial_a  g|_{L_y}=0$ exactly as in \eqref{eq:Tconserved},
\begin{equation}
    \partial^{a_1}T^{(n)}_{a_1a_2\cdots a_n} = 0 .
\end{equation}
These conserved higher-spin currents at the boundary, with the stress tensor at $n=2$, are all built out of the same twistor datum $g$. Note that unlike \(n=2\), these higher-spin charges are frame-dependent since
the denominator marks out the two null directions $q=0,\infty$ of the double-null frame. They also depend on the cohomology representative for $g$; for Dolbeault as above, we would wish to ask that $g$ vanishes near the zeroes and poles of $q$.  

If \v Cech cohomology is used, $g$ is holomorphic on the intersection $U_0\cap U_\infty$ of open sets in an open covering $U_0=\{ |q| < \infty\}$ and $U_\infty= \{ |q|>0\}$ of $\PT$. The integral now is a contour integral over a contour in  $U_0\cap U_\infty$.  Usually in \v Cech cohomology $g$ would have the gauge freedom of adding functions that are holomorphic over all of either $U_0$ or $U_\infty$. These pure gauge contributions will vanish in the original contour integral \eqref{eq:Tpenrose} at spin two, but after multiplying by powers of $q$, the integrals \eqref{eq:higherspinT} will no longer be gauge invariant and will eventually depend on the full function $g$.  This representative can be fixed for example when working globally in split signature, cf  for example \cite{Mason:2022hly} and this corresponds to a choice of boundary conditions in the recursion formulae.

\section{Symmetries}
\label{sec:symmetries}

In this section, following \cite{Bittleston:2024rqe}, we define the $L_\Lambda w_{1+\infty}$ algebra as the algebra of deformations for the complex structure on twistor space. In Section \ref{sec:From complex structure deformations}, we show that these deformations are carried by weight-two holomorphic functions on patch
overlaps. In Section \ref{sec:to metric perturbations}, we work out explicitly the effect of these deformations on the boundary metric. In Section \ref{sec:Structure constants}, we write the structure constants of the $L_\Lambda w_{1+\infty}$ in a monomial basis, discuss the wedge condition, and identify the conformal subalgebra. Here there is an interplay between different  representations of cohomology: the actions are represented in terms of Dolbeault forms, whereas once one has obtained Noether charges, one can use an on-shell formulation with  \v Cech representatives in which the symmetries and cohomology classes are represented in terms of holomorphic functions.

\subsection{From complex structure deformations...}
\label{sec:From complex structure deformations}

The deformations of the complex structure on twistor space, encoded in $h$, are directly related to the normalized contact form $\tau$ introduced earlier via defining the holomorphic volume form for the deformed complex structure to be $\tau \wedge d \tau$.

To deform the standard AdS$_4$ case we can write
\begin{equation}
    \tau = I_{AB}Z^A dZ^B +\frac{1}{\ell^2}h
   \label{tauh} 
\end{equation} where $h\in \Omega^{0,1}(2)$  is a Dolbeault $(0,1)$-form that controls the linearised self-dual perturbations $\overline{\psi}_{\dot\alpha\dot\beta\dot\gamma\dot\delta}$ on spacetime via the Penrose transform with positive helicity. In particular, for pure AdS$_4$, we have $h=0$ as in \eqref{eq:tauinfinity}. As explained in Section \ref{ssec:bulktwistors}, a choice of contact form fixes the conformal class of the boundary metric. Deforming this contact form therefore induces self-dual perturbations of \([g]\). Such deformations generically violate the Dirichlet boundary conditions \eqref{sec:dirichlet falloff} and hence take us outside the original phase space. Similar overleading transformations, or outer symmetries, have appeared in the earlier literature; see, for example, \cite{Campiglia:2016jdj,Campiglia:2016efb,Compere:2017wrj,Compere:2019odm,Compere:2018ylh,Nagy:2022xxs,Nagy:2024dme,Nagy:2024jua}. One possible way to accommodate these transformations would be to consider instead a phase space defined by self-dual boundary conditions, which relate the holographic stress tensor to the boundary Cotton tensor \cite{Mansi:2008bs,Miskovic:2009bm,deHaro:2008gp,Petropoulos:2014yaa}. We leave a detailed investigation of this possibility to future work.

Symmetries correspond to Hamiltonian deformations of $\tau$, since these are the ones that preserve the contact
structure. The Hamiltonian $\xi$ must have  the weight of $\tau$, so $\xi$ is a smooth section of $\cO(2)$.  The Hamiltonian vector field $\{\,\cdot\,,\xi\}$ of the bracket \eqref{eq:poisson} generates a diffeomorphism that deforms $\tau$ by
\begin{equation}
\label{eq:deltatau}
    \delta\tau\big|_y = \frac{1}{\ell^2} \delta h = \frac{1}{\ell^2} \bar\partial\xi .
\end{equation}
To be a symmetry $\dbar \xi=0$ so that $\delta h=0$.  We can consider such a $\xi$ that is a symmetry, hence holomorphic, but  only  locally  as in Section \ref{ssec:penrose} with respect to the cover $U_0=\{\lambda_0\neq0\}$,
$U_1=\{\lambda_1\neq0\}$ of the line, so that the symmetry is a weight-two holomorphic function on
the overlap,
\begin{equation}
\label{eq:symmetrydata}
    \check\xi \ \in\ \cO(2)\!\left(U_0\cap U_1\right) .
\end{equation}
Then such a $\xi$ is ambiguous, it could generate the restriction of a global symmetry, or a diffeomorphism of one of $U_0$ or $U_1$, or a \v Cech representative of $H^1(\PT, \cO(2))$ depending on its singularity structure. 

Unlike the $\cO(-6)$ data of Section \ref{ssec:penrose}, we do not pass to cohomology here:
we consider every $\check\xi$, coboundaries included to incorporate diffeomorphisms in our symmetry algebra since they still act on the
Penrose data through the bracket \eqref{eq:poisson}.

\subsection{... to metric perturbations}
\label{sec:to metric perturbations}
For $\check\xi \in H^1(\PT, \cO(2))$ we can evaluate it as a metric perturbation as follows.  Restrict $\check\xi$ to the line $L_y$ of
a boundary point. Cohomology on the line is trivial in this weight, $H^1(\CP^1,\cO(2))=0$, so the
restriction splits holomorphically,
\begin{equation}
\label{eq:xisplit}
    \check\xi\big|_{L_y} = \xi^{(0)} - \xi^{(1)} ,
\end{equation}
with $\xi^{(i)}$ holomorphic of weight $2$ on the whole of $U_i$. Because $\check\xi$ came from
$\PA$ it is annihilated by the geodesic flow, $p^a \partial_a\check\xi|_{L_y}=0$, so the two halves of
\eqref{eq:xisplit} agree after differentiating,
\begin{equation}
\label{eq:D0xiagree}
    p^a \partial_a\xi^{(0)} = p^a\partial_a\xi^{(1)} .
\end{equation}

Now $p^a \partial_a\xi^{(0)}$ is holomorphic on $U_0$ and, by \eqref{eq:D0xiagree}, equals something
holomorphic on $U_1$; since $U_0\cup U_1 = L_y$ it is holomorphic on the whole line. It has weight
$4$, because $p^a \partial_a$ carries weight $+2$ and $\xi$ carries weight $+2$. A
global holomorphic section of $\cO(4)$ over $L_y$ is a quartic in $\lambda_\alpha$, i.e. a quadratic
in the momentum $p_a\propto\lambda_\alpha\lambda_\beta$, so
\begin{equation}
\label{eq:D0xiisdeltag}
    \boxed{
    \frac{1}{\ell}\,p^a \partial_a\xi^{(i)} = \frac{1}{2}\,\delta g_{ab}\,p^ap^b
    }
\end{equation}
for some symmetric $\delta g_{ab}$ on the boundary. This is the metric perturbation generated by
$[\xi]$.

Two remarks on \eqref{eq:D0xiisdeltag}. First, $H^0(\CP^1,\cO(4))$ is five-dimensional whereas a
symmetric $\delta g_{ab}$ in three dimensions has six components; the one that is missing is the
trace, since $g_{ab}p^ap^b=p^2=0$. The construction is blind to
$\delta g_{ab}\propto g_{ab}$, as it must be, since $\PA$ only ever knew the conformal class
$[g]$. This is the three-dimensional example of the ambitwistor Penrose transform of \cite{Baston:1987av,Mason:2013sva}. Note that unlike the usual Penrose transform, the target fields do not satisfy any field equation. These are general conformal boundary metric perturbations that violate the original Dirichlet boundary conditions \eqref{sec:dirichlet falloff}.

\subsection{Structure constants}
\label{sec:Structure constants}

A representative $\check\xi$ lives on the overlap $U_0\cap U_1$, so it may have
poles where $\lambda_0=0$ or $\lambda_1=0$ but must be regular in $\mu^\alpha$. This gives the basis
of weight-two monomials
\begin{equation}
\label{eq:wmonomials}
    \boxed{w^p_{m,a}
    = \frac{\ell ^{2p+m-a-3}}{(\sqrt{2})^{a-m-2p+5}}\frac{\left(\mu^1\right)^{p+m-1}\left(\mu^0\right)^{a-p+2}}
           {\,\lambda_0^{\,p+a-2}\,\lambda_1^{\,m-p+1}} ,}
\end{equation}
whose weight is $+2$ for every $p,m,a$, the exponents summing to
$(p+m-1)+(a-p+2)-(p+a-2)-(m-p+1)=2$. The requirement that the representatives are regular in $\mu^\alpha$ gives the wedge condition
\begin{equation} \label{wedge condition}
    p+m-1 \geq 0\,, \qquad   a-p +2\geq 0 \, .
\end{equation}
We note that the factor involving $\ell$ is conventional and necessary for a correct flat-space limit when one makes the coordinate transformation to Newman-Unti coordinates $y^\perp = u/\ell, y^- = 2\ell w, y^+ = \bar w/\ell$, where $w,\bar w$ are complex coordinates on $\mathbb{C}$ when appropriate reality conditions are chosen.

 For $\Lambda\neq0$, the infinity twistor is non-degenerate, so
\eqref{eq:tauisPA} gives a Poisson bracket on $\PA$,
\begin{equation}
\label{eq:poisson}
    \{f,h\} = \frac{1}{\sqrt{2}\ell}\frac{\partial f}{\partial\mu^\alpha}\frac{\partial h}{\partial\lambda_\alpha}
            - \frac{1}{\sqrt{2}\ell}\frac{\partial h}{\partial\mu^\alpha}\frac{\partial f}{\partial\lambda_\alpha} ,
\end{equation}
which lowers the total weight by two and so closes on weight-two objects. The monomials
\eqref{eq:wmonomials} therefore inherit an algebra, and hence so do the symmetries they generate,
\begin{equation}
\label{eq:LwAlgebra}
   \boxed{ 
\begin{split}
   \left\{w^p_{m,a},\,w^{p'}_{m',a'}\right\}
    &= \left(m(p'-1)-m'(p-1)\right)w^{p+p'-2}_{m+m',a+a'} \\
    &\qquad\qquad\qquad\qquad\qquad    - \frac{\Lambda}{6}\left(a'(p-2)-a(p'-2)\right)w^{p+p'-1}_{m+m',a+a'} ,
\end{split} 
    }
\end{equation}
the ${L}_\Lambda w_{1+\infty}$ algebra.\footnote{We note that the convention for the cosmological constant differs here compared to \cite{Bittleston:2024rqe}. In this paper, we use the convention that gives a Ricci scalar curvature $R = 4\Lambda$, while in \cite{Bittleston:2024rqe} the convention $R=24\Lambda$ is used. Furthermore, the extra factor of $-1$ can be accounted for by redefining the generators $w^p_{m,a}\rightarrow (-1)^{a-p+2}w^p_{m,a}$.}

Let us show that this infinite-dimensional algebra constitutes an infinite-dimensional enhancement of the conformal algebra \cite{Taylor:2023ajd, Bittleston:2024rqe}. The conformal algebra is the special case in which
$\check\xi|_{L_y}$ has no poles at all, so that it is a global section of $\cO(2)$ over the line.
Such a section is a quadratic in $\lambda_\alpha$, i.e. linear in the momentum,
\begin{equation}
\label{eq:xilinear}
    \check\xi\big|_{L_y} = V_a\,p^a ,
\end{equation}
and then \eqref{eq:D0xiisdeltag} reads $\delta g_{ab}p^ap^b = \frac{2}{\ell}\,\partial_{(a}V_{b)}p^ap^b$. Demanding
that the perturbation vanish, and recalling that only the trace-free part is visible, gives the
conformal Killing equation
\begin{equation}
\label{eq:CKE}
    \partial_{(a}V_{b)} = \frac{1}{3}\,g_{ab}\,\partial_cV^c .
\end{equation}
So the generators with no singularities are exactly the ten conformal Killing vectors of
\eqref{eq:general-boundary-CKV}. Writing a momentum-linear monomial in the affine coordinate as
\begin{equation}
\label{eq:xicomponents}
    w^p_{m,a} = \xi_ap^a
    = \lambda_0^2\left(\tfrac12\xi^- + \xi^\perp q + \tfrac12\xi^+q^2\right) ,
\end{equation}
so that $\xi^-=2[q^0]w^p_{m,a}$, $\xi^\perp=[q^1]w^p_{m,a}$ and $\xi^+=2[q^2]w^p_{m,a}$, exactly ten
of the $w^p_{m,a}$ are of degree $\leq2$ in $q$:
\begin{equation}
\label{eq:conformaltable}
{\renewcommand{\arraystretch}{1.5}
\begin{array}{c|c|c|c|c}
w^p_{m,a}
&
\lambda_0^{-2}w^p_{m,a}
&
(\xi^\perp,\xi^-,\xi^+)
&
\sigma
&
\text{generator}
\\ \hline
w^1_{0,-1}
&
\tfrac12
&
(0,1,0)
&
0
&
P_-
\\
w^{3/2}_{-\frac12,-\frac12}
&
\tfrac12 q
&
(\tfrac12,0,0)
&
0
&
\tfrac12 P_\perp
\\
w^2_{-1,0}
&
\tfrac12 q^2
&
(0,0,1)
&
0
&
P_+
\\
w^1_{0,0}
&
\tfrac{1}{4\ell}(y^-+q y^\perp)
&
\left(
\tfrac{y^\perp}{4\ell},
\tfrac{y^-}{2\ell},
0
\right)
&
\tfrac{1}{4\ell}
&
\tfrac{1}{4\ell}(D-J)
\\
w^{3/2}_{\frac12,-\frac12}
&
\tfrac{\ell}{2}(y^\perp+q y^+)
&
\left(
\tfrac{\ell y^+}{2},
\ell y^\perp,
0
\right)
&
0
&
\tfrac{\ell}{2}B_+
\\
w^{3/2}_{-\frac12,\frac12}
&
\tfrac{1}{4\ell}q(y^-+q y^\perp)
&
\left(
\tfrac{y^-}{4\ell},
0,
\tfrac{y^\perp}{2\ell}
\right)
&
0
&
\tfrac{1}{4\ell}B_-
\\
w^2_{0,0}
&
\tfrac{\ell}{2}q(y^\perp+q y^+)
&
\left(
\tfrac{\ell y^\perp}{2},
0,
\ell y^+
\right)
&
\tfrac{\ell}{2}
&
\tfrac{\ell}{2}(D+J)
\\
w^1_{0,1}
&
\tfrac{1}{8\ell^2}(y^-+q y^\perp)^2
&
\left(
\tfrac{y^\perp y^-}{4\ell^2},
\tfrac{(y^-)^2}{4\ell^2},
\tfrac{(y^\perp)^2}{4\ell^2}
\right)
&
\tfrac{y^-}{4\ell^2}
&
\tfrac{1}{4\ell^2}K_+
\\
w^{3/2}_{\frac12,\frac12}
&
\tfrac14
(y^\perp+q y^+)(y^-+q y^\perp)
&
\left(
\tfrac{(y^\perp)^2+y^-y^+}{4},
\tfrac{y^\perp y^-}{2},
\tfrac{y^\perp y^+}{2}
\right)
&
\tfrac{y^\perp}{2}
&
-\tfrac14 K_\perp
\\
w^2_{1,0}
&
\tfrac{\ell^2}{2}(y^\perp+q y^+)^2
&
\left(
\ell^2y^\perp y^+,
\ell^2(y^\perp)^2,
\ell^2(y^+)^2
\right)
&
\ell^2 y^+
&
\ell^2 K_-
\end{array}}
\end{equation}

Here the generators in the last column are those of \eqref{eq:general-boundary-CKV}, $\sigma$ is the
conformal factor in $\mathcal{L}_\xi[g]=2\sigma[g]$, obtained from
$\sigma=\tfrac13\partial_a\xi^a$, and the explicit column is \eqref{eq:wmonomials} stripped of its
overall $\lambda_0^2$. The two Lorentz entries are the only ones not proportional to a single
generator: $\tfrac12(D\pm J)$ are the two ways of splitting the boost-dilatation pair between the
$q^0$ and $q^2$ ends of \eqref{eq:xicomponents}.

The six generators with integer $p$ form two commuting families,
\begin{equation}
    \left\{w^1_{0,a},w^1_{0,a'}\right\} = \frac{\Lambda}{6}(a'-a)\,w^1_{0,a+a'} ,
    \qquad
    \left\{w^2_{m,0},w^2_{m',0}\right\} = (m-m')\,w^2_{m+m',0} ,
    \qquad
    \left\{w^1_{0,a},w^2_{m,0}\right\} = 0 ,
\end{equation}
i.e. $\mathfrak{sl}(2)\oplus\mathfrak{sl}(2)\cong\mathfrak{so}(2,2)$ acting on the light cone. The
four generators of half-integer weight still give integer powers in the monomial.

Notice that not every element of
\eqref{eq:LwAlgebra} moves the conformal metric. The conformal generators do not, by construction: they
satisfy \eqref{eq:CKE}, so $\delta g_{ab}\propto g_{ab}$ and \eqref{eq:D0xiisdeltag} vanishes. They
act on the boundary as conformal isometries, i.e. as pure diffeomorphisms.

More generally, the trivial ones are visible directly from \eqref{eq:wmonomials}. If a monomial has
poles on one side only --- say $\lambda_0$ appears in the denominator but $\lambda_1$ does not ---
then $\check\xi$ is holomorphic on the whole of $U_0$, so the splitting \eqref{eq:xisplit} can be
taken with $\xi^{(0)}=\check\xi$ and $\xi^{(1)}=0$. Such a $\check\xi$ is a coboundary, and
\eqref{eq:D0xiagree} then forces $p^a\partial_a\xi^{(0)}=0$, so \eqref{eq:D0xiisdeltag} gives
$\delta g_{ab}=0$. The same holds with $\lambda_0$ and $\lambda_1$ exchanged. A monomial therefore
generates a genuine deformation only if it has poles at both $\lambda_0=0$ and $\lambda_1=0$, that
is if
\begin{equation}
\label{eq:polecondition}
    p+a-2 > 0 \qquad\text{and}\qquad m-p+1 > 0 ,
\end{equation}
with $p+m-1\geq0$ and $a-p+2\geq0$ for regularity in $\mu^\alpha$. The ten conformal generators are
those with no poles, where both conditions fail. The remaining generators are of two kinds: those
with poles on one side only are coboundaries that leave the boundary conformal structure unchanged,
while those satisfying both inequalities in \eqref{eq:polecondition} can genuinely shift it.

\section{${L}_\Lambda w_{1+\infty}$ charges and CFT$_3$ light-ray operators}
\label{sec:Charges}

In this section, we derive the ${L}_\Lambda w_{1+\infty}$ charge algebra on twistor space by applying phase space methods on the twistor action. We translate these expressions on spacetime using the Penrose transform assuming linearity. We then relate these expressions to CFT$_3$ light-ray operators at the boundary and recover the ANEC operator and its descendants.

\subsection{Twistor action and phase space methods}

Charge expressions on twistor space have been derived in \cite{Kmec:2024nmu, Kmec:2026dis} starting from a twistor space action \cite{Mason:2007ct} for self-dual gravity. We extend this derivation here for $\Lambda \neq 0$. As for the symmetries above, in this section, we work in the self-dual theory. In the next section, we will translate these results to spacetime in a linearised set-up, without the need to impose a self-duality restriction. 

The fields of the nonsupersymmetric Mason--Wolf theory are $h\in\Omega^{0,1}(\PT,\cO(2))$ and $g\in\Omega^{0,1}(\PT,\cO(-6))$. With the volume convention \eqref{eq:contactnorm},\footnote{Explicitly, the holomorphic top-form in the deformed complex structure $\bar \nabla = \bar\partial+ I^{AB}\partial_A h\partial_B$ is given by $\Omega = \frac{1}{6}\epsilon_{ABCD}Z^A \partial Z^B\wedge \partial Z^C\wedge \partial Z^D$ where the new $(1,0)$-forms are given by $\partial Z^A = dZ^A - I^{BA}\partial_B h$.} where $\tau$ is determined in terms of $h$ via \eqref{tauh}, the twistor action for self-dual gravity reads as
\begin{equation}
\boxed{ S[h,g]=\frac1{2\pi i}\int_{\PT}
       \Omega\wedge g\wedge F_h,
 \qquad F_h=\bar\partial h+\frac12\{h,h\}, }
 \label{eq:MW-action}
\end{equation} where the Poisson bracket $\{ \cdot , \cdot \}$ for $\Lambda \neq 0$ has been defined in \eqref{eq:poisson}. This is the bosonic Poisson-BF action of \cite{Mason:2007ct}, with cosmological constant as in \cite{Bittleston:2024rqe}. Writing $\bar\nabla=\bar\partial+\{h,\cdot\}$, the equations of motion are given by
\begin{equation}
 F_h=0,\qquad \bar\nabla g=0,\qquad
                 \bar\nabla^2=\{F_h,\cdot\}.
 \label{eq:MW-eom}
\end{equation}
The first equation gives an integrable contact complex structure; the second makes $g$ its cotangent field. For a weight $2$ function $\xi$ and a weight $-6$ function $\chi$, the infinitesimal gauge transformations are
\begin{equation}
 \delta_\xi h=\bar\nabla\xi,\qquad
 \delta_\xi g=\{g,\xi\},\qquad
 \delta_\chi h=0,\qquad\delta_\chi g=\bar\nabla\chi.
 \label{eq:deltatau2}
\end{equation}
They preserve \eqref{eq:MW-eom}; invariance of the action follows from the Jacobi identity and $\bar\nabla F_h=0$, up to boundary terms. The $\chi$ transformations with vanishing boundary pairing are quotiented out. In the following, we concentrate on the Hamiltonian transformations generated by $\xi$.

Varying the action and integrating by parts gives, on an oriented real five-dimensional hypersurface $\mathcal C$, the presymplectic potential
\begin{equation}
 \boxed{\Theta_{\mathcal C}[\delta]
   =\frac1{2\pi i}\int_{\mathcal C}
                   \Omega\wedge g\wedge\delta h. }
 \label{eq:MW-potential}
\end{equation}
Its field-space exterior derivative yields the presymplectic form
\begin{equation}
 \omega_{\mathcal C}(\delta_1,\delta_2)
 = \frac1{2\pi i}\int_{\mathcal C} \delta_1 (  \Omega \wedge g) \wedge
 \delta_2h- \delta_2( \Omega \wedge g ) \wedge \delta_1h.
 \label{eq:MW-symplectic}
\end{equation}
Here $\delta_1,\delta_2$ are tangent vectors at an arbitrary solution. Notice that the symplectic structure on twistor space is not affected by the presence of $\Lambda \neq 0$. 

Contracting \eqref{eq:MW-symplectic} with a $\xi$-phase space variation, and following similar steps as in \cite{Kmec:2024nmu, Kmec:2026dis}, we can obtain the Hamiltonians associated with $\xi$. In the following, we evaluate these expressions on-shell; we get a non-integrable charge $\delta H_\xi- \Xi_\xi[\delta]=\omega_{\mathcal C}(\delta,\delta_\xi)$, where 
\begin{equation}
\boxed{  H_\xi=\frac1{2\pi i}
       \int_{\partial\mathcal C}\Omega\wedge\xi g\, , } \qquad \Xi_{\xi}[\delta] = H_{\delta \xi} - \frac{1}{2\pi i}\int_{\partial \mathcal C}\underline{\xi}\lrcorner \, \Omega \wedge g\wedge \delta h \, .
 \label{eq:MW-cut-charge}
\end{equation}
Here $\underline{\xi} = I^{AB} \partial_A \xi \partial_B$ is the Hamiltonian vector field generated by $\xi$. Define the charge bracket by
$\{ H_\xi, H_\eta\}
 :=\delta_\xi H_\eta - \Xi_{\xi}[\delta_\eta]$ \cite{Barnich:2011mi}. On shell, the charge algebra forms a representation of the Poisson bracket
\begin{equation}
 \boxed{ \{ H_\xi, H_\eta\}
             = H_{\{\xi,\eta\}_*}, }
 \label{eq:full-charge-algebra}
\end{equation} where $\{\xi,\eta\}_* = \{\xi,\eta\}+ \delta_\xi \eta - \delta_\eta \xi$ is a modified bracket that takes into account the possible field dependence of the gauge parameters. In our case, $\xi$ will be field independent, $\delta\xi = 0 $.

\subsection{Charge algebra}

In this section, we will give explicit spacetime formulae for the symplectic structure and the ${L}_\Lambda w_{1+\infty}$ charges at the boundary of AdS$_4$. In the bulk, $h$ describes a self-dual non-linear background, $\delta h$ is a self-dual perturbation around it, and $g$ is an anti-self-dual perturbation around it. At the boundary, $h$ encodes the conformal structure, while $g$ controls the holographic stress tensor. In the following, we will evaluate our expressions around $h=0$. Notice that $g$ also induces a linearised perturbation of the boundary Cotton tensor through bulk anti-self-duality \cite{Mansi:2008bs,deHaro:2008gp,Miskovic:2009bm,Petropoulos:2014yaa}. To make the full solution compatible with Dirichlet boundary conditions discussed below \eqref{sec:dirichlet falloff}, we fix the self-dual perturbation $\delta h$ such that the total linearised perturbation of the boundary Cotton tensor, which receives contributions from both $g$ and $\delta h$, vanishes. Hence, $g$ and $\delta h$ generate a linearised bulk Einstein metric on AdS$_4$ obeying Dirichlet boundary conditions, in agreement with the set-up in Section \ref{sec:Linearized gravity in AdS}.

We now wish to prove that the symplectic structure from the twistor action agrees with that for the space-time gravitational degrees of freedom at $\scri$.
The volume form can be written as 
\begin{equation}
    \Omega = \frac{1}{2}\lambda_\alpha\lambda_\beta \d y^{\alpha \gamma}\wedge \d y_\gamma^{\,\,\, \beta}\wedge \D \lambda = p^a \d \Sigma_a \wedge \D \lambda\, , \qquad \d\Sigma_a = \partial_a\lrcorner\,\d^3y\, , \qquad \d^3y = -\frac{1}{2}\,\d y^\perp\wedge\d y^+\wedge\d y^-\, .
\end{equation}
Using \eqref{eq:Tpenrose} and the identity\footnote{This identity follows from using that on the pullback to the spin bundle over $\scri$, $\delta h=\dbar_\lambda \xi $ where $\dbar_\lambda$ is the $\dbar$-operator up the $\CP^1$ fibers because $h$ has weight two and there is no cohomology up the $\CP^1$ fibers for this weight.  Because $p^a\p_a h=0$, $p^a\p_a \xi$ is holomorphic in $p_a$ so $\frac{1}{\ell}p^a\p_a \xi=\frac{1}{2}p^ap^b\delta g_{ab}$ gives a Dolbeault analogue of the discussion following \eqref{eq:xisplit}. 
} $p_a d\Sigma^a \wedge \delta h = \frac{\ell}{2} d^3y \, \delta g_{ab}p^ap^b$ the presymplectic potential \eqref{eq:MW-potential} reads as
\begin{equation}
\boxed{    \Theta[\delta] = \frac{\ell}{2}\int_\mathscr{I}d^3y\,  T^{ab}\delta g_{ab}\, ,}
\end{equation}
where $d^3y$ is the volume form on $\mathscr{I}$.

We can now Penrose transform the twistor expressions for the celestial charges.  For a general generator $\xi$, which is derived from a Čech representative $\check{\xi}$, we define the
associated boundary current
\begin{equation}
   \boxed{ \mathcal J_{\xi a}(y)
    :=
    \frac{1}{2\pi i}
    \int_{L_y}
    \D\lambda\,\wedge 
    p_a\,\xi\,
    g\vert_{L_y}. }
\label{eq:general boundary current}
\end{equation}
The charge \eqref{eq:MW-cut-charge} can then be written as a surface integral
\begin{equation}
    H_\xi[\Sigma]
    =
    \int_\Sigma \d\Sigma^a\,\mathcal J_{\xi a}.
\label{eq:general spacetime charge}
\end{equation}
By using the Čech-Dolbeault isomorphism, one can show that the current $\mathcal J_{\xi a}$ is conserved 
\begin{equation}
\begin{aligned}
    \partial^a\mathcal J_{\xi a}
 =0\, .
\end{aligned}
\label{eq:higher current conservation}
\end{equation}
The full details can be found in Appendix \ref{sec:Charge expressions from Dolbeault to Cech presentations}. Thus, $H_\xi[\Sigma]$ is independent of continuous
deformations of $\Sigma$, and is therefore conserved.  

To make the relation with the celestial charge aspects explicit,
we work in the affine coordinate $q=\lambda_1/\lambda_0$ and define
\begin{equation}
    \widehat{\xi}(q,y)
    :=
    \lambda_0^{-2}\check{\xi}\vert_{L_y}
    =
    \sum_{r\in\mathbb Z}\xi_r(y)q^r,
    \qquad
    \widehat g(q,y)
    :=-
    \lambda_0^6 g\vert_{L_y}.
\end{equation}
In these coordinates,
\begin{equation}
    p_-
    =
    \frac{1}{2}\lambda_0^2,
    \qquad
    p_\perp
    =
    \lambda_0^2q,
    \qquad
    p_+
    =
    \frac{1}{2}\lambda_0^2q^2,
\end{equation}
while the charge aspects are obtained via \eqref{eq:chargeaspecttwistor}, i.e. 
\begin{equation}
    \mathcal Q_s(y)
    =
    \frac{1}{2\pi i}
    \int_{L_y}
    q^{\,s+2}\d q\wedge \widehat g(q,y).
\label{eq:charge aspect moments}
\end{equation}
Choosing the contour orientation consistently with
\eqref{eq:charge aspect moments}, the components of
\eqref{eq:general boundary current} become
\begin{equation}
    \boxed{
    \mathcal J_{\xi+}
    =
    \frac{1}{2}
    \sum_{r\in\mathbb Z}
    \xi_r\,\mathcal Q_r,
    \qquad
    \mathcal J_{\xi\perp}
    =
    \sum_{r\in\mathbb Z}
    \xi_r\,\mathcal Q_{r-1},
    \qquad
    \mathcal J_{\xi-}
    =
    \frac{1}{2}
    \sum_{r\in\mathbb Z}
    \xi_r\,\mathcal Q_{r-2}.
    }
\label{eq:current in terms of charge aspects}
\end{equation}
Consequently, the general spacetime charge reads
\begin{equation}
    \boxed{
    H_\xi[\Sigma]
    =
    \sum_{r\in\mathbb Z}
    \int_\Sigma
    \xi_r(y)
    \left(
        \frac{1}{2}d\Sigma^+\,\mathcal Q_r
        +
        d\Sigma^\perp\,\mathcal Q_{r-1}
        +
        \frac{1}{2}d\Sigma^-\,\mathcal Q_{r-2}
    \right).
    }
\label{eq:general charge aspects charge}
\end{equation}
For a conformal generator, $\widehat{\xi}$ is quadratic in $q$ (see \eqref{eq:xicomponents}),
and \eqref{eq:general charge aspects charge} reduces to
\eqref{eq:conformal-charge-perp-Q}, \eqref{eq:conformal-charge-minus-Q} and \eqref{eq:conformal-charge-plus-Q} after selecting $\Sigma$ to be $\Sigma_\perp$, $\Sigma_-$ and $\Sigma_+$, respectively. For a general
${L}_\Lambda w_{1+\infty}$ generator, poles at $q=0$ or
$q=\infty$ generate moments outside the stress-tensor range
$-2\leq s\leq2$, and hence involve the higher-order charge aspects.

Indeed, restricting the monomial generator \eqref{eq:wmonomials}
to $L_y$ gives
\begin{equation}
\begin{aligned}
    \widehat w^{\,p}_{m,a}(q,y)
    &:=
    \lambda_0^{-2}
    w^{\,p}_{m,a}\vert_{L_y}
    \\
    &=
    \frac{\ell^{\,2p+m-a-3}}{2^{\,a-p+3}}\,
    q^{\,p-m-1}
    \left(y^\perp+qy^+\right)^{p+m-1}
    \left(y^-+qy^\perp\right)^{a-p+2}.
\end{aligned}
\label{eq:w generator on boundary line}
\end{equation}
Under the wedge conditions \eqref{wedge condition}, this is a
finite Laurent polynomial,
\begin{equation}
    \widehat w^{\,p}_{m,a}(q,y)
    =
    \sum_{r=p-m-1}^{p+a}
    w^{\,p}_{m,a;r}(y)\,q^r.
\end{equation}
The charge associated with each basis generator is therefore
\begin{equation}
\boxed{
    H^{\,p}_{m,a}[\Sigma]
    =
    \sum_{r=p-m-1}^{p+a}
    \int_\Sigma
    w^{\,p}_{m,a;r}(y)
    \bigg(
        \frac{1}{2}\d\Sigma^+\,\mathcal Q_{r}
        +\d\Sigma^\perp\,\mathcal Q_{r-1} +\frac{1}{2}\d\Sigma^-\,\mathcal Q_{r-2}
    \bigg).
}
\label{eq:w charges in terms of Qs}
\end{equation}
This provides an explicit spacetime expression for every generator
of the ${L}_\Lambda w_{1+\infty}$ algebra.

A particularly transparent pair of higher-order generators is
\begin{equation}
    \check{\xi}^{(+)}_n
    =
    \lambda_0^2q^{\,n},
    \qquad
    \check{\xi}^{(-)}_n
    =
    \lambda_0^2q^{\,2-n},
    \qquad
    n\geq3.
\end{equation}
Their charges are
\begin{align}
    H_n^{(+)}[\Sigma]
    &=
    \int_\Sigma
    \left(
        \frac{1}{2}d\Sigma^+\,\mathcal Q_n
        +
        d\Sigma^\perp\,\mathcal Q_{n-1}
        +
        \frac{1}{2}d\Sigma^-\,\mathcal Q_{n-2}
    \right),
    \\
    H_n^{(-)}[\Sigma]
    &=
    \int_\Sigma
    \left(
        \frac{1}{2}d\Sigma^+\,\mathcal Q_{2-n}
        +
        d\Sigma^\perp\,\mathcal Q_{1-n}
        +
        \frac{1}{2}d\Sigma^-\,\mathcal Q_{-n}
    \right).
\end{align}
For example, for $n=3$, we have
\begin{align}
    H^{(+)}_3[\Sigma]
    &=
    \int_\Sigma
    \left(
        \frac{1}{2}d\Sigma^+\,\mathcal Q_3
        +
        d\Sigma^\perp\,\mathcal Q_2
        +
        \frac{1}{2}d\Sigma^-\,\mathcal Q_1
    \right),
    \\
    H^{(-)}_3[\Sigma]
    &=
    \int_\Sigma
    \left(
        \frac{1}{2}d\Sigma^+\,\mathcal Q_{-1}
        +
        d\Sigma^\perp\,\mathcal Q_{-2}
        +
        \frac{1}{2}d\Sigma^-\,\mathcal Q_{-3}
    \right).
\end{align}
Thus, a higher-order aspect does not appear in isolation on a
generic hypersurface: it combines with the two adjacent aspects to
form a conserved boundary current. On $\Sigma_-$, for which only
$d\Sigma^+$ is non-vanishing, $H^{(+)}_n$ isolates
$\mathcal Q_n$. Conversely, on $\Sigma_+$, for which only
$d\Sigma^-$ is non-vanishing, $H^{(-)}_n$ isolates
$\mathcal Q_{-n}$.

Equivalently, these charges can be expressed using the conserved
higher-spin currents introduced in Section \ref{sec:Boundary Penrose transform}, see \eqref{eq:higherspinT}. If the restriction of
a generator takes the form
\begin{equation}
    \check{\xi}\vert_{L_y}
    =
    2^{-(n-1)/2}\,
    \Xi^{\alpha_3\cdots\alpha_{2n}}(y)
    \frac{
        \lambda_{\alpha_3}\cdots\lambda_{\alpha_{2n}}
    }{
        \lambda_0^{\,n-2}\lambda_1^{\,n-2}
    },
\end{equation}
then
\begin{equation}
    \mathcal J_{\xi,\alpha_1\alpha_2}
    =
    T^{(n)}_{\alpha_1\cdots\alpha_{2n}}\,
    \Xi^{\alpha_3\cdots\alpha_{2n}},
\end{equation}
or, in tensor notation,
\begin{equation}
   \boxed{ H_\Xi^{(n)}[\Sigma]
    =
    \int_\Sigma
    \d\Sigma^{a_1}\,
    T^{(n)}_{a_1a_2\cdots a_n}\,
    \Xi^{a_2\cdots a_n}. }
\label{eq:higher spin current charge}
\end{equation}
For $n=2$, this reduces to the ordinary conformal charge \eqref{eq:conformal charge} constructed
from the stress tensor. For $n>2$, it gives its higher-spin
generalization and involves the higher-order celestial charge
aspects. The conservation of \eqref{eq:higher spin current charge}
is the spacetime counterpart of the fact that both $\check{\xi}$ and
$g$ are well-defined functions on boundary ambitwistor space.

Notice the striking similarity between \eqref{eq:higher spin current charge} and the quasi-local surface charge integrals written in \cite{Kmec:2026dis} in terms of higher-valence twistors and zero rest mass fields built out of the gravitational field. It would be interesting to make this connection more precise.

\subsection{Light-ray operators in the flat frame}
\label{sec:light-ray-operators-flat}

From the previous subsection, we learned that every generator
$\check\xi$ defines a conserved boundary current
$\mathcal J_{\xi a}$ and an associated surface charge,
\begin{equation}
    \partial^a\mathcal J_{\xi a}=0,
    \qquad
    H_\xi[\Sigma]
    =
    \int_\Sigma d\Sigma^a\,\mathcal J_{\xi a}.
\label{eq:conserved-current-and-charge}
\end{equation}
A light-ray operator is obtained by integrating one component of
this current along a complete null generator. For the two null
congruences generated by $\partial_+$ and $\partial_-$, respectively,
we define
\begin{align}
    \mathcal L_{\xi,+}(y^\perp,y^-)
    &:=
    \int_{-\infty}^{+\infty}dy^+\,
    \mathcal J_{\xi+}(y^\perp,y^+,y^-),
    \label{eq:plus-light-transform}
    \\
    \mathcal L_{\xi,-}(y^\perp,y^+)
    &:=
    \int_{-\infty}^{+\infty}dy^-\,
    \mathcal J_{\xi-}(y^\perp,y^+,y^-).
    \label{eq:minus-light-transform}
\end{align}
Thus, the light-ray operators are simply the charge densities with
respect to the transverse coordinate $y^\perp$. Indeed, recall that
\begin{equation}
    \Sigma_-:=\{y^-=\mathrm{const}\},
    \qquad
    \Sigma_+:=\{y^+=\mathrm{const}\}.
\end{equation}
With the orientations used in Section \ref{sec:Celestial charge aspects and conformal symmetries}, the corresponding charges
can be written as
\begin{align}
    H_\xi[\Sigma_-]
    &=
    \int dy^\perp dy^+\,
    \mathcal J_{\xi+}
    =
    \int dy^\perp\,
    \mathcal L_{\xi,+}(y^\perp,y^-),
    \label{eq:charge-plus-light-transform}
    \\
    H_\xi[\Sigma_+]
    &=
    \int dy^\perp dy^-\,
    \mathcal J_{\xi-}
    =
    \int dy^\perp\,
    \mathcal L_{\xi,-}(y^\perp,y^+).
    \label{eq:charge-minus-light-transform}
\end{align}
Consequently, choosing a null integration surface does not define a
new charge: it gives a light-ray representation of the same conserved
charge constructed in the previous subsection. 

For a conformal generator, the restriction of $\check\xi$ to the
ambitwistor line takes the form
\begin{equation}
    \left.\check\xi\right|_{L_y}
    =
    \zeta^a p_a,
\end{equation}
where $\zeta^a$ is a conformal Killing vector. The associated current
then reduces to the usual stress-tensor current,
\begin{equation}
    \mathcal J_{\xi a}
    =
    T_{ab}\zeta^b.
\label{eq:conformal-current}
\end{equation}
In particular, for the translation generated by
$\zeta=\partial_+$, the light transform
\eqref{eq:plus-light-transform} is the averaged null energy operator (ANEC)
\begin{equation}
    \mathcal E(y^\perp)
    =
    \int_{-\infty}^{+\infty}dy^+\,
    T_{++}(y^\perp,y^+,y^-=0).
\label{eq:ANEC-flat-complex}
\end{equation}

The coordinates $(y^\perp,y^+,y^-)$ used in the preceding sections
describe either the complexified boundary or a Klein-signature real
slice, for which $y^+$ and $y^-$ are independent null coordinates.
To recover the physical Lorentzian operator while preserving the
null direction $\partial_+$, we choose the real slice
\begin{equation}
    y^+=\gamma^+,
    \qquad
    y^-=-\gamma^-,
    \qquad
    y^\perp=-i\gamma^\perp,
    \qquad
    \gamma^\perp,\gamma^\pm\in\mathbb R.
\label{eq:complex-to-Lorentzian-slice}
\end{equation}
The complexified boundary metric then restricts to
\begin{equation}
    -(dy^\perp)^2+dy^+dy^-
    =
    -d\gamma^+d\gamma^-+(d\gamma^\perp)^2
    =
    ds_{\mathrm M_3}^2, 
\end{equation}
which corresponds to an analytic continuation to Lorentzian signature. In particular,
$\partial_{y^+}=\partial_{\gamma^+}$ and
$T^{(\mathbb C)}_{++}=T^{(\mathrm M)}_{++}$ after analytic
continuation. Hence
\begin{equation}
    \boxed{\mathcal E_{\mathrm M}(\gamma^\perp)
    :=
    \int_{-\infty}^{+\infty}d\gamma^+\,
    T^{(\mathrm M)}_{++}
    (\gamma^\perp,\gamma^+,0)
    =
    \left.
    \mathcal E(y^\perp)
    \right|_{y^\perp=-i\gamma^\perp}. }
\label{eq:ANEC-Lorentzian-slice}
\end{equation}
By footnote~\ref{fn:stress-normalization}, $\int d\gamma^+\langle\mathcal T_{++}\rangle=-\frac{\ell}{4\pi G_{\mathrm N}}\mathcal E_{\mathrm M}$, so in the present normalization the positivity of ANEC reads
$\mathcal E_{\mathrm M}(\gamma^\perp)\leq0$. Its expression in the complexified
$y$-coordinates should be understood as its analytic continuation.
An analogous discussion applies to the light transforms along
$\partial_-$.

\subsection{Light-ray operators on the Einstein cylinder}
\label{sec:light-ray-operators-cylinder}

We now express the same light-ray operators in the
Einstein-cylinder conformal frame to compare with the analysis of \cite{Sheta:2025oep,Strominger:2026yrh}. Reference \cite{Sheta:2025oep} considers the
light transform of a dimension-two conserved CFT current. It first
works in the boundary metric inherited from global AdS$_4$ and then
Weyl-rescales this metric to the round Einstein cylinder. On the
null surface, this rescaling converts the unweighted current integral
into an integral with measure
$d\tau^+\sin\tau^+$.
Reference \cite{Strominger:2026yrh} applies the same construction to
the stress tensor currents and relates the resulting operators
directly to their flat space expressions. We use the coordinates and
the direct conformal map of the latter reference.

Let $(\gamma^\perp,\gamma^+,\gamma^-)$ denote the Lorentzian flat coordinates introduced
above and let $(\tau^+,\tau^-,\varphi)$ denote the coordinates on the
Einstein cylinder. The conformal map is
\begin{equation}
\begin{aligned}
    \gamma^+
    &=
    -2\Omega\cos\tau^+\cos\tau^-,
    \\
    \gamma^-
    &=
    2\Omega\sin\tau^+\sin\tau^-,
    \\
    \gamma^\perp
    &=
    \Omega\sin(\tau^+-\tau^-)\sin\varphi,
\end{aligned}
\qquad
    \Omega
    =
    \frac{1}{
        \sin(\tau^++\tau^-)
        +
        \sin(\tau^+-\tau^-)\cos\varphi
    }.
\label{eq:Minkowski-to-cylinder-map}
\end{equation}
It obeys
\begin{equation}
    ds_{\mathrm M_3}^2
    =
    \Omega^2 ds_{\mathrm{EC}_3}^2,
\end{equation}
where
\begin{equation}
    ds_{\mathrm{EC}_3}^2
    =
    -4\,d\tau^+d\tau^-
    +
    \sin^2(\tau^+-\tau^-)\,d\varphi^2,
    \qquad
    \tau^\pm=\frac{1}{2}(\tau\pm\theta).
\label{eq:Einstein-cylinder-metric}
\end{equation}
Since there is no Weyl anomaly in three dimensions, the covariant
stress tensor transforms as
\begin{equation}
    T^{(\mathrm M)}_{\mu\nu}(\gamma)
    =
    \Omega^{-1}
    \frac{\partial X^A}{\partial \gamma^\mu}
    \frac{\partial X^B}{\partial \gamma^\nu}
    T^{(\mathrm{EC})}_{AB}(X(\gamma)),
    \qquad
    X^A=(\tau^+,\tau^-,\varphi).
\label{eq:stress-tensor-Weyl-map}
\end{equation}

The null surface relevant for the light transform is
\begin{equation}
    \Sigma_{\mathrm{null}}
    =
    \left\{
        \tau^-=0,\quad
        0\leq\tau^+\leq\pi,\quad
        0\leq\varphi<2\pi
    \right\}.
\label{eq:null-cylinder-surface}
\end{equation}
It is the portion of the future light cone joining a point of the
Einstein cylinder to its antipode. The curves at fixed $\varphi$ are
its null generators, while their union forms a null Cauchy surface.

On $\Sigma_{\mathrm{null}}$, the conformal map reduces to
\begin{equation}
    \gamma^-=0,
    \qquad
    \gamma^\perp=\tan\frac{\varphi}{2},
    \qquad
    \gamma^+=-(1+(\gamma^\perp)^2)\cot\tau^+,
    \qquad
    \Omega=\frac{1+(\gamma^\perp)^2}{2\sin\tau^+}.
\label{eq:null-surface-coordinate-map}
\end{equation}
In terms of the complex coordinates used in the preceding
subsection, this becomes
\begin{equation}
    y^-=0,
    \qquad
    y^\perp=-i\tan\frac{\varphi}{2},
    \qquad
    y^+=-
    \left(
        1+\tan^2\frac{\varphi}{2}
    \right)
    \cot\tau^+.
\label{eq:null-surface-complex-map}
\end{equation}
Thus, the complete Lorentzian null line
$-\infty<\gamma^+<+\infty$ at fixed $\gamma^\perp$ is mapped to the finite null
segment $0<\tau^+<\pi$ at fixed $\varphi$. Equivalently, it is the
Lorentzian real slice of the complex null line at
$y^\perp=-i\gamma^\perp$.

For a general conserved current, the conformal transformation is
most conveniently formulated directly in terms of the flux two-form
entering the charge. If
$F:\mathrm{EC}_3\rightarrow\mathrm M_3$ denotes the map
\eqref{eq:Minkowski-to-cylinder-map}, we define the current in the
cylinder frame by
\begin{equation}
    \star_{\mathrm{EC}}
    \mathcal J_\xi^{(\mathrm{EC})}
    =
    F^*
    \left(
        \star_{\mathrm M}
        \mathcal J_\xi^{(\mathrm M)}
    \right).
\label{eq:current-flux-transport}
\end{equation}
This avoids assigning a separate Weyl weight to every component of
$\mathcal J_{\xi a}$. With the orientation used above, its restriction
to the null surface is
\begin{equation}
    \left.
    \star_{\mathrm{EC}}
    \mathcal J_\xi^{(\mathrm{EC})}
    \right|_{\Sigma_{\mathrm{null}}}
    =
    d\tau^+\wedge d\varphi\,
    \sin\tau^+\,
    \mathcal J_{\xi+}^{(\mathrm{EC})}
    (\tau^+,0,\varphi).
\label{eq:current-flux-null-surface}
\end{equation}
Since $dy^\perp=-i\,d\gamma^\perp$, the charge on the null Cauchy surface therefore takes the form
\begin{equation}
    H_\xi[\Sigma_{\mathrm{null}}]
    =
    -i\int_0^{2\pi}d\varphi\,
    \mathcal L_\xi^{\mathrm{LR}}(\varphi),
\label{eq:charge-as-light-transform}
\end{equation}
where
\begin{equation}
    \mathcal L_\xi^{\mathrm{LR}}(\varphi)
    :=
    \int_0^\pi d\tau^+\,
    \sin\tau^+\,
    \mathcal J_{\xi+}^{(\mathrm{EC})}
    (\tau^+,0,\varphi).
\label{eq:general-light-transform}
\end{equation}
This is the current light transform of
\cite{Sheta:2025oep}, written in the normalization adopted here.
For a conformal generator, using
$\mathcal J_{\xi\mu}=T_{\mu\nu}\zeta^\nu$, it also gives the
stress-tensor light transforms considered in
\cite{Strominger:2026yrh}.

Let us now specialize to the ANEC operator. Along a null generator,
\begin{equation}
    \frac{d\gamma^+}{d\tau^+}
    =
    (1+(\gamma^\perp)^2)\csc^2\tau^+,
    \qquad
    \frac{d\varphi}{d\gamma^\perp}
    =
    \frac{2}{1+(\gamma^\perp)^2}.
\label{eq:null-line-Jacobians}
\end{equation}
Combining the first relation with
\eqref{eq:stress-tensor-Weyl-map} gives
\begin{equation}
    d\gamma^+\,
    T^{(\mathrm M)}_{++}
    =
    \frac{2\sin^3\tau^+}{(1+(\gamma^\perp)^2)^2}\,
    d\tau^+\,
    T^{(\mathrm{EC})}_{++}.
\label{eq:ANEC-integrand-transformation}
\end{equation}
It follows that
\begin{equation}
    \mathcal E_{\mathrm M}(\gamma^\perp)
    =
    \left(
        \frac{d\varphi}{d\gamma^\perp}
    \right)^2
    \mathcal E_{\mathrm{EC}}(\varphi),
    \qquad
    \gamma^\perp=\tan\frac{\varphi}{2},
\label{eq:ANEC-frame-transformation}
\end{equation}
where
\begin{equation}
    \mathcal E_{\mathrm{EC}}(\varphi)
    =
    \frac{1}{2}
    \int_0^\pi d\tau^+\,
    \sin^3\tau^+\,
    T^{(\mathrm{EC})}_{++}(\tau^+,0,\varphi).
\label{eq:ANEC-cylinder}
\end{equation}
The flat and cylinder expressions therefore describe the same
light-ray operator. The factor in
\eqref{eq:ANEC-frame-transformation} reflects the fact that they are
densities with respect to different coordinates, $\gamma^\perp$ and
$\varphi$, on the space of null generators.

The natural modes on the cylinder are
\begin{equation}
    \mathcal E_k
    =
    \int_0^{2\pi}d\varphi\,
    e^{ik\varphi}\,
    \mathcal E_{\mathrm{EC}}(\varphi),
    \qquad
    k\in\mathbb Z.
\label{eq:ANEC-modes}
\end{equation}
Using
\begin{equation}
    e^{i\varphi}
    =
    \frac{1+i\gamma^\perp}{1-i\gamma^\perp},
    \qquad
    d\varphi=\frac{2\,d\gamma^\perp}{1+(\gamma^\perp)^2},
\end{equation}
together with \eqref{eq:ANEC-frame-transformation}, these modes can
equivalently be written in the Lorentzian flat frame as
\begin{equation}
    \mathcal E_k
    =
    \int_{-\infty}^{+\infty}d\gamma^\perp\,
    \frac{1+(\gamma^\perp)^2}{2}
    \left(
        \frac{1+i\gamma^\perp}{1-i \gamma^\perp}
    \right)^k
    \mathcal E_{\mathrm M}(\gamma^\perp).
\label{eq:ANEC-modes-flat-frame}
\end{equation}

To make contact with \cite{Strominger:2026yrh}, we perform a residual conformal transformation, which acts on twistor space by the linear change of coordinates
\begin{equation}
    \lambda'_0=\mu^0+\frac{1}{\sqrt{2}}\lambda_1,
    \qquad
    \lambda'_1=-\mu^0+\frac{1}{\sqrt{2}}\lambda_1,
    \qquad
    \mu'^0=\frac{1}{\sqrt{2}}\mu^1-\frac{1}{2}\lambda_0,
    \qquad
    \mu'^1=\frac{1}{\sqrt{2}}\mu^1+\frac{1}{2}\lambda_0 .
\label{eq:primed-twistor-coordinates}
\end{equation}
This transformation places the light rays generating $\Sigma_{\mathrm{null}}$ at $\lambda'_\alpha=0$, and it preserves the bracket \eqref{eq:poisson}:
\begin{equation}
    \{\mu'^\alpha,\lambda'_\beta\}=\frac{1}{\sqrt{2}\ell}\,\delta^\alpha_\beta,
    \qquad
    \{\mu'^\alpha,\mu'^\beta\}=\{\lambda'_\alpha,\lambda'_\beta\}=0 .
\end{equation}
Consequently the primed monomials
\begin{equation}
    w'^p_{m,a}
    = \frac{\ell^{2p+m-a-3}}{(\sqrt{2})^{a-m-2p+5}}\,\frac{\left(\mu'^1\right)^{p+m-1}\left(\mu'^0\right)^{a-p+2}}{(\lambda'_0)^{p+a-2}\,(\lambda'_1)^{m-p+1}}
\label{eq:primed-monomials}
\end{equation}
are a different set of twistor functions that obey \eqref{eq:LwAlgebra} with the same structure constants: they realize the same $L_\Lambda w_{1+\infty}$ algebra. We denote with a prime the other quantities of Sections~\ref{sec:Asymptotically AdS spacetimes}--\ref{sec:Charges} constructed from $(\mu'^\alpha,\lambda'_\alpha)$ in place of $(\mu^\alpha,\lambda_\alpha)$: the boundary coordinates $y'^a$, defined by $\mu'^\alpha=y'^{\alpha\beta}\lambda'_\beta$ and given by
\begin{equation}
    y'^-=\frac{y^+y^--(1-y^\perp)^2}{2y^-},
    \qquad
    y'^\perp=\frac{y^+y^-+1-(y^\perp)^2}{2y^-},
    \qquad
    y'^+=\frac{y^+y^--(1+y^\perp)^2}{2y^-};
\end{equation}
the affine coordinate $q'=\lambda'_1/\lambda'_0$; the charge aspects $\mathcal Q'_s$; the Fefferman--Graham coefficients $\Psi'^{(m)}_{n,\mathrm{FG}}$ in the Poincar\'e patch with boundary coordinates $y'^a$; and the charges $H'$.

We now identify these modes with the primed generators lying on the edge of
the $L_\Lambda w_{1+\infty}$ wedge. Reference \cite{Strominger:2026yrh} denotes the generators by
$w^p_{\bar m,m}$. To distinguish their mode labels from ours, we
write them as $(\bar m_{\mathrm{CFT}},m_{\mathrm{CFT}})$. The
relation between the two conventions is
\begin{equation}
    \bar m_{\mathrm{CFT}}=m,
    \qquad
    m_{\mathrm{CFT}}=a.
\label{eq:CFT-mode-identification}
\end{equation}
Accordingly, the wedge conditions of
\cite{Strominger:2026yrh},
\begin{equation}
    \bar m_{\mathrm{CFT}}+p\geq1,
    \qquad
    m_{\mathrm{CFT}}-p\geq-2,
\end{equation}
coincide precisely with the conditions
\eqref{wedge condition}. The ANEC modes lie on the edge obtained
by saturating both inequalities. In the conventions of
\cite{Strominger:2026yrh}, they are
\begin{equation}
    \mathcal E_k
    =
    -i^k
    w^{\frac{3+k}{2}}_
      {-\frac{k+1}{2},\,\frac{k-1}{2}}.
\label{eq:ANEC-w-generator}
\end{equation}
Setting
\begin{equation}
    p=\frac{k+3}{2},
    \qquad
    m=1-p=-\frac{k+1}{2},
    \qquad
    a=p-2=\frac{k-1}{2},
\label{eq:ANEC-edge-labels}
\end{equation}
the restriction of the corresponding primed generator to $L_{y'}$ simplifies
to
\begin{equation}
    \widehat w'^{\,p}_{1-p,p-2}(q',y')
    =
    \frac{1}{2}q'^{\,2p-2}
    =
    \frac{1}{2}q'^{\,k+1}.
\label{eq:ANEC-twistor-generator}
\end{equation}

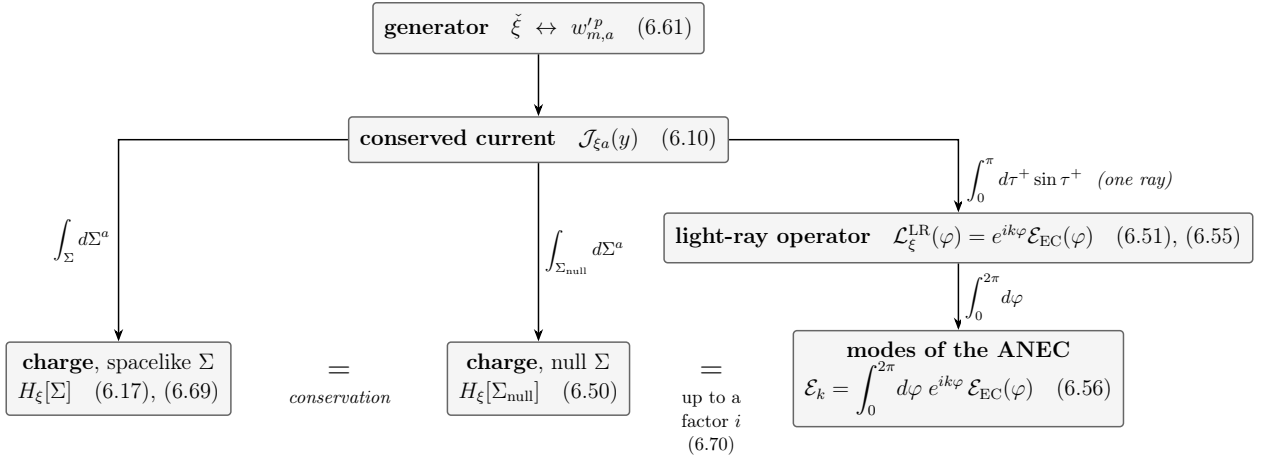
\begin{figure}[t]\centering
\resizebox{\textwidth}{!}{%
\begin{tikzpicture}[
  box/.style={draw=black!60, rounded corners=2pt, inner sep=6pt, align=center, fill=black!4, thick},
  lab/.style={font=\footnotesize\itshape, midway, fill=white, inner sep=2pt},
  arr/.style={->, >=Stealth, thick},
  every node/.style={align=center}]
\node[box] (gen) at (0, 0) {\textbf{generator}\quad $\check\xi\ \leftrightarrow\ w'^{\,p}_{m,a}$\quad \eqref{eq:primed-monomials}};
\node[box] (cur) at (0, -2.1) {\textbf{conserved current}\quad $\mathcal J_{\xi a}(y)$\quad \eqref{eq:general boundary current}};
\node[box] (lro) at (7.9, -3.95) {\textbf{light-ray operator}\quad
  $\mathcal L^{\mathrm{LR}}_\xi(\varphi)=e^{ik\varphi}\mathcal E_{\mathrm{EC}}(\varphi)$\quad \eqref{eq:general-light-transform}, \eqref{eq:ANEC-cylinder}};
\node[box] (asp) at (-7.9, -6.6) {\textbf{charge}, spacelike $\Sigma$\\[1pt]
  $\displaystyle H_\xi[\Sigma]$\quad \eqref{eq:general charge aspects charge}, \eqref{eq:ANEC-charge-aspects}};
\node[box] (nul) at (0, -6.6) {\textbf{charge}, null $\Sigma$\\[1pt]
  $\displaystyle H_\xi[\Sigma_{\mathrm{null}}]$\quad \eqref{eq:charge-as-light-transform}};
\node[box] (mod) at (7.9, -6.6) {\textbf{modes of the ANEC}\\[1pt]
  $\displaystyle \mathcal E_k=\int_0^{2\pi}\!d\varphi\; e^{ik\varphi}\,\mathcal E_{\mathrm{EC}}(\varphi)$\quad \eqref{eq:ANEC-modes}};
\draw[arr] (gen) -- (cur);
\draw[arr] (cur.west) -- ++(-0.4,0) -| node[lab, pos=0.75, left=2pt]
  {$\displaystyle\int_{\Sigma}d\Sigma^a$} (asp.north);
\draw[arr] (cur.south) -- node[lab, right=1pt]
  {$\displaystyle\int_{\Sigma_{\mathrm{null}}}d\Sigma^a$} (nul.north);
\draw[arr] (cur.east) -- ++(0.4,0) -| node[lab, pos=0.78, right=2pt]
  {$\displaystyle\int_0^\pi d\tau^+\sin\tau^+$\ \ (one ray)} (lro.north);
\draw[arr] (lro.south) -- node[lab, right=1pt]
  {$\displaystyle\int_0^{2\pi}\!d\varphi$} (mod.north);
\node[align=center, fill=white, inner sep=2pt, anchor=north] at ($(asp.east)!0.5!(nul.west)+(0,0.3)$)
  {\Large$=$\\[-1pt]{\footnotesize\itshape conservation}};
\node[align=center, fill=white, inner sep=2pt, anchor=north] at ($(nul.east)!0.5!(mod.west)+(0,0.3)$)
  {\Large$=$\\[-1pt]{\footnotesize up to a}\\[-2pt]{\footnotesize factor $i$}\\[-2pt]{\footnotesize \eqref{eq:ANEC-charge-identification}}};
\end{tikzpicture}}%
\caption{The quantities of this section and the maps between them. Three ways of arriving at the same spacetime charge corresponding to the primed edge generators
$\widehat w'=\tfrac12q'^{k+1}$.}
\label{fig:quantity-map}
\end{figure}

To express the corresponding spacetime charge in terms of the
primed celestial charge aspects, let us denote by
$\mathcal J'^{[r]}_a$ the conserved current associated with the
monomial $\widehat\xi'=q'^r$. From the general current formula
\eqref{eq:current in terms of charge aspects}, its components are
\begin{equation}
    \mathcal J'^{[r]}_+
    =
    \frac{1}{2}\mathcal Q'_r,
    \qquad
    \mathcal J'^{[r]}_\perp
    =
    \mathcal Q'_{r-1},
    \qquad
    \mathcal J'^{[r]}_-
    =
    \frac{1}{2}\mathcal Q'_{r-2}.
\label{eq:monomial-current}
\end{equation}
The conservation of this current is equivalent to the recursion
relation \eqref{HigherRecursionRelations}. The charge associated
with the ANEC edge generator is therefore
\begin{equation}
\begin{split}
    H'^{\mathrm{edge}}_k[\Sigma]
    &:=
    H'^{\frac{3+k}{2}}_
      {-\frac{k+1}{2},\,\frac{k-1}{2}}[\Sigma]
    =
    \frac{1}{2}
    \int_\Sigma d\Sigma'^a\,
    \mathcal J'^{[k+1]}_a
    \\
    &=
    \int_\Sigma
    \left(
        \frac{1}{4}d\Sigma'^+\,\mathcal Q'_{k+1}
        +
        \frac{1}{2}d\Sigma'^\perp\,\mathcal Q'_k
        +
        \frac{1}{4}d\Sigma'^-\,\mathcal Q'_{k-1}
    \right).
\end{split}
\label{eq:ANEC-charge-aspects}
\end{equation}
On $y^-=0$ the primed edge generator restricts to
$\frac{1}{2}\big(1-(y^\perp)^2\big)\big(\frac{1-y^\perp}{1+y^\perp}\big)^kp_+$,
which is the smearing of the ANEC modes in
\eqref{eq:ANEC-modes-flat-frame} at $y^\perp=-i\gamma^\perp$. Since
$dy^\perp=-i\,d\gamma^\perp$, the identification between the ANEC
modes and the spacetime charges is
\begin{equation}
    \boxed{
    \mathcal E_k
    =
    i\,
    H'^{\mathrm{edge}}_k[\Sigma_{\mathrm{null}}]
    =
    i\,
    H'^{\mathrm{edge}}_k[\Sigma].
    }
\label{eq:ANEC-charge-identification}
\end{equation}
Here $\Sigma$ is any Cauchy surface that can be continuously
deformed to $\Sigma_{\mathrm{null}}$. Thus, on
$\Sigma_{\mathrm{null}}$ the charge is represented as an ANEC
light transform, whereas on a different Cauchy surface it is
expressed in terms of the primed celestial charge aspects. This structure
is summarized in Figure~\ref{fig:quantity-map}.

For $k=-1,0,1$, all three charge aspects appearing in
\eqref{eq:ANEC-charge-aspects} lie in the stress-tensor range
$-2\leq s\leq2$. These three modes are proportional to the three primed
translation charges. The first modes involving aspects outside this
range are
\begin{align}
    H'^{\mathrm{edge}}_{-2}[\Sigma]
    &=
    \int_\Sigma
    \left(
        \frac{1}{4}d\Sigma'^+\,\mathcal Q'_{-1}
        +
        \frac{1}{2}d\Sigma'^\perp\,\mathcal Q'_{-2}
        +
        \frac{1}{4}d\Sigma'^-\,\mathcal Q'_{-3}
    \right),
    \\
    H'^{\mathrm{edge}}_{2}[\Sigma]
    &=
    \int_\Sigma
    \left(
        \frac{1}{4}d\Sigma'^+\,\mathcal Q'_3
        +
        \frac{1}{2}d\Sigma'^\perp\,\mathcal Q'_2
        +
        \frac{1}{4}d\Sigma'^-\,\mathcal Q'_1
    \right).
\end{align}
They are one half of the charges associated with
$\widehat\xi'=q'^{-1}$ and $\widehat\xi'=q'^3$, respectively. Using the
results of Section~\ref{sec:Higher-order celestial charge hierarchy},
the additional charge aspects are
\begin{align}
    \mathcal Q'_3
    &=
    \partial_-'^{-1}
    \left(
        \partial'_+\mathcal Q'_1
        -
        \frac{1}{\ell}\Psi'^{(1)}_{0,\mathrm{FG}}
    \right),
    \\
    \mathcal Q'_{-3}
    &=
    \partial_+'^{-1}
    \left(
        \partial'_-\mathcal Q'_{-1}
        +
        \frac{1}{\ell}\Psi'^{(1)}_{4,\mathrm{FG}}
    \right).
\end{align}
Thus, the first non-conformal ANEC modes already involve the first
higher-order celestial charge aspects.

The same construction extends from the ANEC edge to the full
$L_\Lambda w_{1+\infty}$ wedge. Every point in the wedge
can be parametrized as
\begin{equation}
    m=1-p+u,
    \qquad
    a=p-2+v,
    \qquad
    u,v\in\mathbb Z_{\geq0}.
\label{eq:wedge-descendant-parametrization}
\end{equation}
Writing again $k=2p-3$, the corresponding generator takes the form
\begin{equation}
\begin{split}
    \widehat w'_{k;u,v}(q',y')
    &:=
    \widehat w'^{\,p}_{1-p+u,p-2+v}(q',y')
    \\
    &=
    \frac{\ell^{u-v}}{2^{v+1}}\,
    q'^{\,k+1-u}
    \left(y'^\perp+q'y'^+\right)^u
    \left(y'^-+q'y'^\perp\right)^v.
\end{split}
\label{eq:general-ANEC-descendant-generator}
\end{equation}
The ANEC modes correspond to $u=v=0$, while increasing $u$ or $v$
moves into the interior of the wedge. Expanding the two polynomials
and using \eqref{eq:monomial-current}, one obtains the following
uniform expression for the corresponding charge:
\begin{equation}
\begin{split}
    H'_{k;u,v}[\Sigma]
    &=
    \frac{\ell^{u-v}}{2^{v+1}}
    \sum_{i=0}^{u}\sum_{j=0}^{v}
    \binom{u}{i}\binom{v}{j}
    \int_\Sigma d\Sigma'^a\,
    (y'^\perp)^{u-i+j}
    (y'^+)^i
    (y'^-)^{v-j}
    \mathcal J'^{[k+1-u+i+j]}_a .
\end{split}
\label{eq:general-ANEC-descendant-charge}
\end{equation}
For fixed $(k,u,v)$, this expression involves only finitely many
celestial charge aspects, whose labels satisfy
\begin{equation}
    k-1-u
    \leq s\leq
    k+1+v.
\end{equation}
For example, the first two directions away from the ANEC edge give
\begin{align}
    H'_{k;1,0}[\Sigma]
    &=
    \frac{\ell}{2}
    \int_\Sigma d\Sigma'^a
    \left(
        y'^\perp\mathcal J'^{[k]}_a
        +
        y'^+\mathcal J'^{[k+1]}_a
    \right),
    \\
    H'_{k;0,1}[\Sigma]
    &=
    \frac{1}{4\ell}
    \int_\Sigma d\Sigma'^a
    \left(
        y'^-\mathcal J'^{[k+1]}_a
        +
        y'^\perp\mathcal J'^{[k+2]}_a
    \right).
\end{align}
In particular, the generalized Cordova--Shao operators of
\cite{Strominger:2026yrh} are
\begin{align}
    \mathcal N_k+i\mathcal K_k
    &=
    -4\ell\,
    H'_{k-1;0,1}[\Sigma],
    \\
    \mathcal N_k-i\mathcal K_k
    &=
    \frac{2}{\ell}\,
    H'_{k+1;1,0}[\Sigma].
\end{align}

Reference~\cite{Strominger:2026yrh} gives explicit expressions for
the ANEC operator, the generalized operators $\mathcal K$ and
$\mathcal N$, and the light transforms of the ten conformal
currents. The remaining elements of the wedge are generated
algebraically by repeated conformal transformations and
commutators. Equation
\eqref{eq:general-ANEC-descendant-charge} complements this
construction by providing a uniform spacetime representative for
every such descendant in terms of the primed celestial charge aspects
$\mathcal Q'_s$.

The tensors introduced in \eqref{eq:higherspinT} should
not, in a generic interacting CFT$_3$, be interpreted as additional
independent local higher-spin primary currents. Rather, in the
linearised construction considered here, they provide a convenient
spacetime packaging of non-local light-ray descendants of the stress
tensor. The inverse-derivative prescriptions entering the definition
of the higher $\mathcal Q_s$ correspond to choices of boundary
conditions and endpoint prescriptions for these non-local
operators. Finally, the identification above is semiclassical: the
twistor Poisson algebra gives the classical limit of the quantum
commutator algebra established in
\cite{Strominger:2026yrh}.

\section{Outlook}

In this work, we have derived the ${L}_\Lambda w_{1+\infty}$ charges on twistor space in the presence of a non-vanishing cosmological constant by applying phase-space methods to the twistor action for (non-linear) self-dual gravity.  At the conformal boundary, the bulk twistor space is identified with the complexification of the space of null geodesics in the boundary, the boundary ambitwistor space, via the LeBrun Heaven on Earth construction. 
The charges are those that generate local holomorphic contact diffeomorphisms of twistor space (diffeomorphisms that preserve the contact structure defined by the `infinity twistor') and are expressed in terms of twistor data that encodes the full gravitational phase space at $\scri$ including both the conformal structure of $\scri$ and the holographic stress-energy tensor.
Using the appropriate boundary Penrose transform, we have translated these charges into spacetime at $\scri$ in the limit when the boundary conformal structure is trivial
as is natural when using Dirichlet boundary conditions and as is usual for the study of CFTs on conformally flat backgrounds.
The celestial algebra of Poisson diffeomorphisms of twistor space can be built from the standard geometric symmetries of $\scri$ by an algebraic recursion; when realized on space-time at $\scri$ this becomes an integro-differential recursion for generating the full hierarchy of charges.   We have then interpreted the resulting charges in terms of light-ray operators in the dual $\mathrm{CFT}_3$.

These results open several directions for future investigation. First, an important direction is to extend our construction beyond the conformally flat background on spacetime at $\scri$. A priori the Heaven on Earth construction generates a self-dual bulk AdS$_4$ metric but this cannot in general be guaranteed to be global without special assumptions  \cite{Biquard:2002} and the relationship of such fillings with those in conventional holography is not clear. In particular, it would be valuable to recover, within the ambitwistor framework, the expressions recently obtained from a spacetime analysis \cite{DiGiacomo:2026oku}, in which self-duality conditions must be imposed.  More generally one should use a twistor action for full Einstein gravity. This won't siginificantly change Noether arguments at $\scri$, but becomes important in the bulk. Unlike the Yang-Mills case, the status of the twistor action for full Einstein gravity is not clear, although see \cite{Adamo:2013tja} for a route into the subject that restricts the twistor action for full conformal gravity \cite{Mason:2005zm} to Einstein modes.  It would also be interesting to clarify the relation with the earlier $\Lambda$-BMS framework \cite{Compere:2019bua, Compere:2020lrt, Fiorucci:2020xto}, which describes cosmological-constant deformations of generalized BMS symmetries in de Sitter and anti-de Sitter spacetimes.

A closely related direction is to apply our methods to Yang--Mills theory in $\mathrm{AdS}_4$, with the aim of recovering the $\mathrm{CFT}_3$ light-ray operators discussed in \cite{Sheta:2025oep}, which realize the $S$-algebra. This setting is particularly promising for two reasons. First, a formulation of self-dual Yang--Mills theory in $\mathrm{AdS}_4$ was recently given in \cite{Heuveline:2026knw}, providing a natural framework in which one may expect to construct the complete set of $S$-algebra generators. Second, since a twistor action for full Yang--Mills theory is known \cite{Mason:2005zm}, it may be possible to develop a phase-space construction of these charges beyond the self-dual sector.

Another important question concerns the Ward identities associated with these charges. Their conservation and interpretation as symmetry generators suggest an infinite family of Ward identities extending the usual conformal Ward identities. It would be important to establish their precise form and to determine whether they persist beyond the linearised or self-dual sectors. In self-dual Yang--Mills theory, such identities may help explain the remarkable simplicity of the correlation functions found in \cite{Heuveline:2026knw}. More generally, an important question is as to whether there are any examples of CFT$_3$s that admit the full ${L}_\Lambda w_{1+\infty}$ algebra as full symmetries or just approximate, in particular if there are nontrivial such theories admitting the full symmetries that are interacting. Such theories must therefore have a life on ambitwistor space if they admit the full set of symmetries.  One possiblity is that such theories might have to be completely integrable, but on the other hand, the ambitwistor correspondence does not classically imply field equations so integrability is not implied and beyond such integrable sectors, these algebras might still provide structural information about more general CFTs.

Finally, it would be interesting to investigate the flat-space limit of our results. Bondi coordinates, which admit a natural generalization to $\mathrm{AdS}_4$ \cite{Poole:2018koa, Compere:2019bua, Compere:2020lrt, Ruzziconi:2019pzd, Campoleoni:2023fug, Geiller:2022vto}, should provide a convenient framework for this purpose. Taking the flat-space limit in the bulk is expected to induce a Carrollian limit of the boundary theory \cite{Ruzziconi:2026bix}, and we expect the ${L}_\Lambda w_{1+\infty}$ generators identified here to reduce to the soft operators at null infinity, which could themselves be interpreted as extended/non-local operators \cite{Ruzziconi:2026isv} in a Carrollian CFT.

\section*{Acknowledgments}

We warmly thank Atul Sharma for discussions at an early stage of the project. We are also grateful to Simon Heuveline, Ana-Maria Raclariu, Ahmed Sheta and Andy Strominger for useful exchanges. 

 AG and LM are supported by the Simons Foundation Collaboration on Celestial Holography. LM and AK are supported by the STFC on ST/T000864/1. RR is supported by the European Union’s Horizon Europe research and innovation programme under the Marie Sklodowska-Curie grant agreement No. 101104845 (UniFlatHolo), hosted at Harvard University and École Polytechnique. Preprint
number: CPHT-RR024.082026. OpenAI and Anthropic LLMs were used at several stages of
this project. All results were author-verified.

\appendix

\section{Further twistor details}\label{Appendix: Further Details}

In this appendix, we provide further details on the twistor theory sections presented in the main text.

\subsection{Bulk twistor space and Lax pair}
\label{sec:Bulk twistor space and Lax pair}

Another way of characterising twistor space \eqref{def twistor} which we will make use of is as a quotient of the spin bundle.
Take the projectivised spin bundle $\PS = M\times\CP^1_\lambda$, whose points are a spacetime point
together with a choice of $\sigma_\alpha$ up to scale.
On it define the Lax pair
\begin{equation}
\label{eq:laxpair}
    L_{\dot\alpha} = \sigma^\alpha e_{\alpha\dot\alpha}  - \Gamma_{\alpha\dot\alpha\beta}^{\quad\,\,\gamma} \sigma^\alpha \sigma^\beta \frac{\partial}{\partial \sigma^\gamma},
    \qquad \dot\alpha = \dot0,\dot1 .
\end{equation}
where $e_{\alpha\dot\alpha}$ is a tetrad basis. Any combination $c^{\dot\alpha}L_{\dot\alpha}$ moves in the direction
$c^{\dot\alpha}\sigma^\beta$, which is again rank one and hence null, so the two vector fields
span exactly the $\beta$-plane picked out by $\sigma_\alpha$. The pair commutes, $[L_{\dot0},L_{\dot1}] = 0$ mod $L_{\dot\alpha}$ when the background is self-dual $\psi_{\alpha\beta\gamma\delta} = 0$, so the
distribution they span is integrable and its leaves are the $\beta$-surfaces. The space of leaves is twistor space,
\begin{equation}
\label{eq:PTquotient}
    \PT = \PS\big/\left\langle L_{\dot0},L_{\dot1}\right\rangle . 
\end{equation}
The incidence relation \eqref{eq:incidence} is the statement that $\mu^{\dot\alpha}$ and
$\lambda_\alpha$ are constant along the leaves,
\begin{equation}
    L_{\dot\alpha}\mu^{\dot\beta} = L_{\dot\alpha}\lambda_\beta = 0 ,
\end{equation}
so they are coordinates on the quotient. In this language, a twistor function is simply a
function on the spin bundle annihilated by the Lax pair. Explicitly, for the metric \eqref{Poincare metric}, and writing $\partial_\pm = \partial/\partial y^\pm$, the Lax operators are equivalent to
\begin{equation}
\label{eq:laxcoords}
    \sqrt{2}L_{\dot 0} = -2\sigma_1\partial_- + \sigma_0(\partial_\perp-\partial_z) -\frac{\sigma_0}{2z} \Upsilon_{\mathbb{CP}^1}\, ,
    \quad
    \sqrt{2}L_{\dot 1} = -\sigma_1(\partial_\perp + \partial_z)+2\sigma_0\partial_+  - \frac{\sigma_1}{2z}\Upsilon_{\mathbb{CP}^1} \, ,
\end{equation}
where $\Upsilon_{\mathbb{CP}^1} = \sigma_\alpha\frac{\partial}{\partial \sigma_\alpha}$ is the Euler vector field on $\mathbb{CP}^1$.  
The solutions to the incidence relations give rise to an important distinction between the twistor space coordinate $\lambda_\alpha$ and the spin bundle coordinate over spacetime $\sigma_\alpha$. The explicit incidence relations in the bulk are given by \begin{equation} \label{sigma coord}
    \mu^{\dot \alpha} = x^{\dot \alpha\alpha}\lambda_{\alpha}\, , \quad  \lambda_\alpha = \left(\frac{\ell}{z}\right)^{\frac{1}{2}} \sigma_\alpha
\end{equation} such that the contact structure $\tau_{L_x} = \D \sigma$ restricts to the holomorphic 1-form on the spin bundle. Therefore, the incidence relations on AdS are the same as those for Minkowski space rescaled by a factor of $\left(\frac{\ell}{z}\right)^{\frac{1}{2}} $.

\subsection{More on ambitwistor space}
\label{sec:More on ambitwistor space}

In this section, we provide more details of the ambitwistor space, which is the space of complexified null geodesics of the conformal boundary metric. Formally, 
we take the cotangent bundle, restrict to null directions, and
quotient by the geodesic flow,
\begin{equation}
\label{eq:ambidef}
    \PA = T^*_N N\big/D_0 , \qquad
    T^*_N N = \left\{(y,p)\in T^*N \,\big|\, g^{-1}(p,p) = 0\right\} ,
\end{equation}
the flow being generated by
\begin{equation}
\label{eq:D0gen}
    D_0 = p^a\left(\frac{\partial}{\partial y^a}
      + \Gamma^c_{ab}p_c\frac{\partial}{\partial p_b}\right) ,
\end{equation}
The cotangent bundle comes equipped with a natural symplectic
structure $\omega = \d p_a\wedge\d y^a$. With respect to \(\omega\), \(D_0\) is the
Hamiltonian vector field of $H = -\tfrac12 g^{ab}p_ap_b$ so that \(\mathbb{PA}\) is the symplectic reduction of the cotangent bundle by the constraint \(p^2 = 0\). This means that it comes with a natural one-form, the symplectic potential $\vartheta=p_a\,dy^a$, which satisfies
on the null cone
\begin{equation} \label{eq:taudescends}
    \mathcal{L}_{D_0}\vartheta=0
    \qquad
    D_0\mathbin{\lrcorner}\vartheta=0.
\end{equation}
and therefore descends to the quotient \cite{LeBrun:1982vjh}. In the affine coordinate $q=\lambda_1/\lambda_0$, the geodesic
generator for the metric \eqref{conf boundary metric} reads
\begin{equation}
\label{eq:D0coords}
    D_0=\lambda_0^2
    \left(\partial_+-q\,\partial_\perp+q^2\,\partial_-\right).
\end{equation}

 We had two separate constructions of \(\mathbb{PA} \cong \mathbb{PT}\) that both involved taking quotients. It is worth relating the two explicitly. The crux of the identification is that from the bulk, this space is the space of totally null two-planes. These meet the
boundary in its null geodesics, which is our other characterisation of the same space; see Figure \ref{fig:heaven-on-earth}. The
Lax pair spans these 2-planes while $D_0$ spans the boundary null geodesics. i.e., \(D_0\) spans one direction of the totally null 2-planes of the bulk.

Explicitly, $D_0$ is a certain linear combination of the Lax-pair elements,
\begin{equation}
\label{eq:D0fromlax}
D_0
= \frac{\ell}{\sqrt{2}\,z}
\,\sigma_\alpha N^{\alpha\dot\alpha}L_{\dot\alpha}
= \lambda_0^2
\left(\partial_+-q\,\partial_\perp+q^2\partial_-\right)
\end{equation}
Here, the vector field $N^{\alpha\dot\alpha}$ is given by
\begin{equation}
N^{\alpha\dot\alpha}
= 
\begin{pmatrix}
0 & 1\\
-1 & 0
\end{pmatrix}
\end{equation}
and corresponds to the normalized normal
$N=\frac{\sqrt{2}\,z}{\ell}\,\partial_z$, satisfying $N^2=2$.

\subsection{Proof of the Penrose transform in AdS}
\label{sec:Proof of the Penrose transform in AdS}

Let us show that the integral expression \eqref{eq:penrosespin2} solves \eqref{eq:zrm2}. The field equation comes from the Lax pair condition of Section
\ref{sec:Bulk twistor space and Lax pair}. By using the incidence relations \eqref{sigma coord} and the homogeneous weight of $g$, the integral formula \eqref{eq:penrosespin2} can be written as
\begin{equation}
    \psi_{\alpha\beta\gamma\delta} = \frac{z}{\ell}\frac{1}{2\pi i}\oint_{\Gamma}\D\sigma \, \sigma_\alpha\sigma_\beta\sigma_\gamma \sigma_\delta \, g(x\cdot\sigma, \sigma)\, .
\end{equation}
We recognise that the resulting contour integral satisfies the flat z.r.m. equation such that 
\begin{equation}
     \psi_{\alpha\beta\gamma\delta} = \frac{z}{\ell}  \psi_{\alpha\beta\gamma\delta}^{\text{flat}}\, ,\qquad \partial^{\alpha\dot\alpha} \psi_{\alpha\beta\gamma\delta}^{\text{flat}} = 0\,.
\end{equation}
The z.r.m. equation is covariant under a Weyl rescaling, provided the metric scales as $\hat g = \Omega^2 g^{\text{flat}}$ and the z.r.m. field scales as $ \psi_{\alpha\beta\gamma\delta} = \Omega^{-1}\psi_{\alpha\beta\gamma\delta}^{\text{flat}}$. The transformation of the z.r.m. equation under Weyl rescaling is given by
\begin{equation}
    \nabla^{\alpha\dot\alpha}\psi_{\alpha\beta\gamma\delta} = \Omega^{-3}\partial^{\alpha\dot\alpha}\psi_{\alpha\beta\gamma\delta}^{\text{flat}}\, .
\end{equation}
In the particular case of $\Omega = \frac{\ell}{z}$, we find that $\psi_{\alpha\beta\gamma\delta} = \frac{z}{\ell}\psi_{\alpha\beta\gamma\delta}^{\text{flat}}$ satisfies the z.r.m. equation on an AdS$_4$ background.

We note that the integral formula \eqref{eq:penrosespin2} naturally lives in a spin frame where the Lax operator is equivalent to $L_{\dot\alpha} = \lambda^\alpha \partial_{\alpha\dot\alpha}$ where the partial derivative keeps $\lambda_\alpha$ constant. This is a spin frame where the ASD spin bundle has been made trivial, but the SD spin bundle still has a connection. To go to the physical spin frame, we must multiply each undotted index by a factor of $(\frac{z}{\ell})^{1/2}$. This transformation to the physical spin bundle then gives the correct fall-off for the z.r.m. field used in the rest of the text.

\subsection{Charge expressions from Dolbeault to \v{C}ech presentations}
\label{sec:Charge expressions from Dolbeault to Cech presentations}

In \cite{Kmec:2024nmu, Kmec:2026dis}, phase space methods have been applied on the twistor action for self-dual gravity to obtain Noether's charges. In particular, this construction also holds for linearised gravity, which is the focus of the present work. The formulation of the twistor action and the application of phase space methods requires the Dolbeault formulation of twistor theory, which allows for an off-shell formulation of twistor space data. In this section, we rewrite the charge expression of \cite{Kmec:2024nmu, Kmec:2026dis} in a  \v{C}ech presentation to match with the main text. 

We perform the explicit Čech-Dolbeault isomorphism; to do this, begin by considering a partition of unity $\{f_0,f_1\}$ subordinate to the cover $\{U_0,U_1\}$ of $\mathbb{CP}^1$. This means that $f_i$ only has support in $U_i$ and the condition $f_0+f_1=1$ holds. To go from the Čech representative $g_{01}$ defined on the overlap $U_0\cap U_1$ to a global Dolbeault representative $g$ we define $g_0 = g_{01}f_1$ and $g_1 = -g_{01}f_0$ which have support on $U_1$ and $U_0$ respectively. These have the property that on $U_0\cap U_1$ we have 
\begin{equation}
    g_0 - g_1 = g_{01}(f_0+f_1) = g_{01}\, ,
\end{equation}
which is holomorphic on the overlap. Therefore, on the overlap
\begin{equation}
    \bar \partial g_0 - \bar\partial g_1 = \bar\partial g_{01} = 0
\end{equation}
since $g_{01}$ is holomorphic on $U_0\cap U_1$. We find that the $\bar\partial g_i$ agree on overlaps, and as a consequence, give a globally defined $(0,1)$-form $g$, which is naturally closed, and is defined patch-wise $g\vert_{U_i} = \bar\partial g_i $. Let $\xi$ be any global extension agreeing with $\check{\xi}$ on the support of $g$, which is contained in the overlap $U_0\cap U_1$. The current inside the charge integral then becomes
\begin{equation}
\begin{split}
    \int_{\mathbb{CP}^1} \D \lambda \wedge p_a\xi g\vert_y &= \int_{U_0\cap U_1} \D\lambda \wedge p_a \check{\xi} g_{01} \bar \partial f_1 \\ &= \oint_\Gamma \D\lambda p_a\check \xi g_{01}\, ,
\end{split}
\end{equation}
where the first line uses the explicit representative for $g$ which only has support on the overlap due to $\bar\partial f_1$, and going from the first to second line, we use Stokes' theorem and the fact that both $\check{\xi}$ and $g_{01}$ are holomorphic on the overlap.

We will use these results to translate the Dolbeault charge expression to their Čech versions, making their translation to spacetime more direct. In particular, the general boundary current \eqref{eq:general boundary current} is given by 
\begin{equation}
    \mathcal J_{\xi a}(y)
    :=
    \frac{1}{2\pi i}
    \oint_{\Gamma}
    \D\lambda\,
    p_a\,\check{\xi}\,
    g_{01}\vert_{L_y}. 
\label{eq:general boundary current_app}
\end{equation}
Since both $\check{\xi}$ and $g$ descend from ambitwistor space and are
therefore constant along the boundary null-geodesic flow,
\begin{equation}
    p^a\partial_a\check{\xi}=0,
    \qquad
    p^a\partial_a g_{01}\vert_{L_y}=0.
\end{equation}
It follows immediately that
\begin{equation}
    \partial^a \mathcal J_{\xi a} =
    \frac{1}{2\pi i}
    \oint_\Gamma
    \D\lambda\,
    \left[
        \left(p^a\partial_a\check{\xi}\right)g_{01}
        +
        \check{\xi}\left(p^a\partial_a g_{01}\right)
    \right]_{L_y}
    =0.
\end{equation}

\addcontentsline{toc}{section}{References}
\bibliographystyle{style}
\bibliography{references}

\end{document}